\documentclass[twocolumn,showpacs,amsmath,amssymb,amsfonts,superscriptaddress,aps,prstper,10pt]{revtex4-1}
\usepackage{amsmath}
\usepackage{graphics}% Include figure files
\usepackage{dcolumn}% Align table columns on decimal point
\usepackage{bm}% bold math
\usepackage{enumerate}
\usepackage{ctable}
\usepackage{booktabs}
\usepackage{natbib}
\newcommand{\ER}{Erd\"{o}s-Renyi\ }
\newcommand{\mytab}{\phantom{XXXX}}
\newcommand{\mycom}[1]{{\bfseries\itshape{\#~#1}}}

\begin{document}
\title{An empirical approach to interpreting card-sorting data}
\author{Steven F.~Wolf}
\email{wolfste4@msu.edu}
\affiliation{%
Department of Physics and Astronomy
Michigan State University, East Lansing, MI 48824, USA
}%
\affiliation{%
Lyman Briggs College 
Michigan State University, East Lansing, MI 48825, USA
}%
\author{Daniel P.~Dougherty}
\email{doughe57@msu.edu}
\affiliation{%
Lyman Briggs College 
Michigan State University, East Lansing, MI 48825, USA
}%
\author{Gerd Kortemeyer}
\email{korte@lite.msu.edu}
\affiliation{%
Department of Physics and Astronomy
Michigan State University, East Lansing, MI 48824, USA
}%
\affiliation{%
Lyman Briggs College 
Michigan State University, East Lansing, MI 48825, USA
}%
\date{\today}
\begin{abstract}
Since it was first published 30 years ago, \citeauthor{Chi81}'s seminal paper on expert and novice categorization of introductory problems led to a plethora of follow-up studies within and outside of the area of physics [\citeauthor{Chi81} Cognitive Science \textbf{5}, 121 -- 152 (1981)].  These studies frequently encompass ``card-sorting'' exercises whereby the participants group problems.  While this technique certainly allows insights into problem solving approaches, simple descriptive statistics more often than not fail to find significant differences between experts and novices. In moving beyond descriptive statistics, we describe a novel microscopic approach that takes into account the individual identity of the cards and uses graph theory and models to visualize, analyze, and interpreting problem categorization experiments.  We apply these methods to an introductory physics (mechanics) problem categorization experiment, and find that most of the variation in sorting outcome is not due to the sorter being an expert versus a novice, but rather due to an independent characteristic that we named ``stacker'' versus ``spreader.'' The fact that the expert-novice distinction only accounts for a smaller amount of the variation may explain the frequent null-results when conducting these experiments.\end{abstract}
\maketitle

\section{Introduction}
Physics education is key to the development of new physicists and to the development of the field through physics research.  Other than working in research labs doing fundamental physics research, many physicists go on to solve important problems while working in other fields.  An effective physics curriculum clearly prepares students for a diverse array of jobs, which explains in part why physicists are well-known for being resourceful problem solvers.  \citeauthor{Lark80} concluded that the basis of this problem solving ability is the array of cognitive connections between multiple concepts, making each physics concept a part of a coherent whole rather than disparate bits of information~\cite{Lark80}. Fuller points out the importance of a good conceptual understanding when he says, ``Every physicist knows the importance of having the correct concept in mind \textit{before} beginning to solve a problem,''~\cite[emphasis ours]{Fuller82}.

Categorization studies comparing experts and novices started with \citeauthor{Chi81}, who studied the categorization of introductory physics problems~\cite{Chi81}.  This study, which, to date, has been cited over 3000 times, has been critical in the study of the differences between experts and novices in many areas such as Clinical Psychology~\cite{MurphWright84}, dinosaur expertise~\cite{JohnScoMerv04}, wine tasting~\cite{Solomon97}, and even Star Wars philosophy~\cite{StarWars}.  All of these studies go back to the same apparently straightforward result in physics: novices categorize introductory physics problems by ``surface features'' (e.g.~ ``incline,'' ``pendulum,'' or ``projectile motion''), while experts use ``deep structure'' (e.g.~ ``energy conservation'' or ``Newton's second law'').  We now take a closer look at previous categorization studies involving physics problems.

\section{Previous Card-Sorting Studies}
The novice group of \citeauthor{Chi81}'s study was made up of eight students who had just finished the first semester of an introductory university physics class, and the expert group was made up of eight advanced Ph.D.~physics students.  Both groups were given the instructions to sort the problems ``based on similarity of solution''\cite{Chi81}.  Problems were allowed to be placed in two (or more) categories if the sorter so desired; we call this ``multiple categorization,'' as opposed to ``single categorization,'' where each problem would have to be sorted in one and only one category.

Each sorter categorized their set in front of a member of the research team according to a uniform protocol.  Sorters were required to sort the problems without paper and pencil to prevent them from actually solving the problems.  After sorting the problems a second time --- to check for consistency --- the sorters explained the reasoning for their groupings.  After a qualitative analysis of the category names used by more than two sorters, \citeauthor{Chi81}'s group concluded that the key distinction between experts and novices is, quite sensibly, that experts sort problems based on the physics principle required to solve each problem, while novices sort the problems based on surface features.  This difference in categorization, \citeauthor{Chi81} concluded, was an experts' ability to convert contextual cues from the problem texts and figures into the physics principles that are required to solve those problems.  The main message from Chi's paper is that this difference in categorization behavior allows experts to be better problem solvers than novices \cite{Chi81}.

In a subsequent study, Veldhuis attempted to verify the result of \citeauthor{Chi81} \cite{veld90}.  Veldhuis had three groups, a novice group comprised of 94 introductory physics students, an intermediate group of 5 students who had just finished classical mechanics, and an expert group of 20 physics professors---among whom only 2 had not taught calculus-based physics.  Veldhuis created four different categorization sets, one of which was given to each subject to categorize according to a protocol similar to that used by \citeauthor{Chi81}'s group.  The first set was created in an attempt to mimic the \citeauthor{Chi81} problem set \cite{veld_thesis}, and the second was a control set with a similar collection of end-of-chapter problems. In contrast, the third and fourth sets were carefully constructed so that each problem had only a single physics principle and a single surface feature from a set of principles and surface features \cite{veld90}.  For example, Table \ref{matrix} shows how the third set was constructed by populating a matrix of four surface and four conceptual features. The fourth set was also ``rigged.''  It had the same number of cards, but only two surface and two conceptual features. \citeauthor{veld_thesis} could not draw a conclusion from the categorizations from his first two problem sets.  However, sets 3 and 4 agreed with \citeauthor{Chi81} in that experts categorize problems based on physics principles while novices show a ``more complex behavior.'' \cite{veld90,veld_thesis}  Ironically, \citeauthor{veld90} observed that distinguishing experts and novices based on surface features of their categorizations failed unless the desired physics features --- conceptual and surface --- were built into the design of the experiment.

\begin{table}
\caption{\textbf{\citeauthor{veld90}'s Matrix Method.}  Deep Structures are listed along the top.  Surface Features are listed along the left.  By ``terms'' \citeauthor{veld90} includes ``physical arrangements of objects and literal physics terms'' in the problem text.\cite{veld_thesis}  \citeauthor{veld90} created this set hoping that experts would group the problems by column and novices would group the problems by row.
\label{matrix}}
\begin{ruledtabular}
\begin{tabular}{lllll}
				&Newton II\footnote{Newton's second law}		&E cons\footnote{Energy conservation}		&$\vec{p}$ cons\footnote{Linear momentum conservation}	&$\vec{L}$ cons\footnote{Angular momentum conservation}\\ \hline \\[0em]
Spring	&Prob 16	 		&Prob 2		&Prob 4					&Prob 9\\
Ramp		&Prob 11 			&Prob 6		&Prob 12				&Prob 15\\
Pulley	&Prob 5	 			&Prob 14	&Prob 13				&Prob 8\\
Terms		&Prob 3	 			&Prob 10	&Prob 7					&Prob 1\\
\end{tabular}
\end{ruledtabular}
\end{table}

More recently, the work done in Singh's group at the University of Pittsburgh has broadened the application of ``card-sorting'' to other fields~\cite{Mason09,Mason11,Singh09,Lin10}.  \citeauthor{Mason11}  compared students in introductory physics courses with both physics graduate students and physics faculty. \citeauthor{Mason11} created two categorization sets of twenty-four problems each.  The first set  was created in an attempt to mimic \citeauthor{Chi81}'s set. Seven problems were directly from \citeauthor{Chi81}'s original set, based on examples given in the paper, while the remainder of the \citeauthor{Chi81}'s original set is apparently lost in history.  A second set was devised because the results from the first set showed ``major differences'' with \citeauthor{Chi81}'s data~\cite{Mason09,Mason11}, which may not be surprising given Veldhuis's previous results~\cite{veld90,veld_thesis}.  Each subject, upon reading the problems, filled in three columns on a response sheet:  category name, the appropriateness of the category name, and the identity of problems that fit in the category.  \citeauthor{Mason11} then rated each problem's category as ``good,'' ``moderate,'' or ``poor'' based on each sorter's description of the category.  A category was considered `good' if it was based on the underlying physics principles. He then asked a faculty panel to validate his ratings by following the same procedure on a subset of the categorizations.  

\citeauthor{Mason11} found that the problems taken directly from \citeauthor{Chi81}'s original study were placed by novices in ``good'' categories far less often than they did on average, determining that they were generally from topics more difficult to novice students.  For example, difficult topics for novices might have been rotational motion, non-equilibrium applications of Newton's 2nd law, or the Work-Energy theorem \cite{Mason09,Mason11}.  \citeauthor{Mason11} also found that the superficial category names were far less prevalent in his study than in Chi's original study.  It is possible that the shift away from novices' use of superficial category names is due to a change in curricular focus precipitated by \citeauthor{Chi81}'s result.  Contrary to the sharp distinction found by \citeauthor{Chi81}, \citeauthor{Mason11} found that there was some overlap between the calculus-based introductory physics students and the graduate students~\cite{Mason09,Mason11}.  

In a follow-up study, \citeauthor{Singh09}~\cite{Singh09} asked graduate student teaching assistants to perform a similar categorization exercise, both as themselves and through the eyes of their students, and compared both types of their categorizations  to physics faculty and introductory students. In contrast with \citeauthor{Chi81}, \citeauthor{Singh09} considered the physics faculty as the ``true experts'' and only looked at graduate students as a sort of intermediate group. Similar to \citeauthor{Mason11}, problem categories were rated to be ``good'', ``moderate,'' or ``poor,'' validated by a faculty panel.  \citeauthor{Singh09} found that the graduate students acting as introductory students performed better on the categorization task than did actual introductory students, thus overestimating their students~\cite{Singh09}.  Singh found that the professors performed best on the categorization task, distinguishing this group from the categorizations of the graduate students acting as themselves.  This suggested that the use of graduate students as an expert group is not entirely accurate, as their behavior is not truly expert-like.

Finally, in a separate study, \citeauthor{Lin10} also carried out a categorization study concerning Quantum Mechanics problems \cite{Lin10}.  For this task the novice group consisted of twenty-two Junior and Senior physics majors taking Quantum Mechanics.  The expert group consisted of six faculty members \cite{Lin10}.  In contrast to the previous studies mentioned here, \citeauthor{Lin10} chose to have a three-member faculty panel evaluate all of the categorizations, scoring each category as either good, moderate, or poor.  In contrast to the studies of introductory physics problems, in \citeauthor{Lin10}'s study, the expert group had more variability, as even the faculty panel did not see this task in stark terms.  Two of the panel members even said that they disliked using the terms ``good'' and ``poor'' to describe a categorization of Quantum Mechanics problems; this reservation was not voiced by the raters in the introductory problem categorization studies \cite{Lin10}.  Similarly, the faculty panel members said that sometimes they preferred another categorization choice to their own \cite{Lin10}.  All of this, \citeauthor{Lin10} conclude, was due to the more difficult nature of the problems.  In any case, it is clear that no ``ideal'' set of groupings existed, and it was impossible to simply assign some ``score'' to a given categorization. 

In summary, replicating \citeauthor{Chi81}'s seminal experiment is challenging. More often than not, attempts to repeat it fail, as an informal survey among physics education researchers indicates --- however, such null-results do not get published. Yet, as a community of physics educators, we hold a firm belief that deep down there is a significant difference in problem solving behavior between experts and novices, and that categorization is an important piece of the puzzle. Quantifying this difference, however, more often than not, remains elusive.

\section{Method Philosophy}
While \citeauthor{Chi81}'s method has been the predominant paradigm for follow-up studies, their methodology is based on a certain model of the categorization process.  Using a different model, one will arrive at a different methodology. Given the importance of this experimental technique, we believe it is important to understand the underlying model and consider alternatives to its assumptions.

\subsection{Macroscopic versus Microscopic Cluster Comparison}
 \citeauthor{Chi81}'s group looked at a processed version of the category names agreed upon by multiple sorters and counted the number of problems in each category name \cite{Chi81}.  Their analysis does not seem to hinge on the identity of the problems in each group, merely the number of problems in that group.  For example, if two sorters both used the category name ``Conservation of Energy'' but one sorter put problems $\left\{1,3,5,7,9\right\}$ in that set and the other sorter put problems $\left\{2,4,7,8,9\right\}$ in that set, \citeauthor{Chi81}'s analysis would count that as two people who both used an energy related variant as a category and both had five problems in that set.  In other words, the sets would be treated identically.  We argue that it is important that these two groups should be treated \emph{differently}, as they have few identical elements. We believe that instead of just these ``macroscopic'' measures (sizes and names of groups), the sorting results should also be compared on the ``microscopic'' level of individual problems.
 
\subsection{Deterministic versus Variable Nature of Sorting}
Different understandings of the underlying process of categorization will lead to different statistical analysis methods.  \citeauthor{Chi81} seem to view categorization as a deterministic process, as evidenced by the ``double-check'' step in their experimental method.  They see any minor replication variation as evidence of an underlying method. On the other hand, one of the phenomena that physics education research has to grapple with is the variability of learner responses to what appear to be identical scenarios, see for example \citeauthor{Frank08}~\cite{Frank08}. Rather than interpreting card-sorting outcomes as reflections of stable theories or beliefs, an alternative model is that they are based on {\it ad hoc} assemblies of more simple intuitions (similar to ``phenomenological primitives,''~\cite{diSessa93} or ``resources''~\cite{hammer00}) --- those are then assembled ``on-the-fly,'' and the particular assembly may vary on circumstances. There is no reason to expect that card-sorting experiments are immune to this variability, and one may thus expect that any sorter who categorizes the same set of problems on separate occasions would return different results, although he or she might even recognize the problems that are used.  We cannot control the actual mechanisms potentially underlying these ``random'' outcomes, but have accounted for the resulting variability in the choice of our statistical methods. In addition, we use sample-based statistics to interpret our categorization data, realizing that our sample is only part of a vastly larger population.  

\subsection{Parametric versus Non-Parametric Scoring}
Previous analysis methods \citep{Mason09,Mason11,Singh09,Lin10} describe each categorization individually with a score, which is either a comparison to an ``ideal'' categorization set or an individual ``grade'' of each set.  These methods measure performance on the categorization task, where the scoring criteria is an input of the evaluation process --- the process starts with assumptions of what properties an expert categorization will have.  It may, however, not be clear what an ``ideal'' set is, which in turn makes the scoring somewhat ambiguous. First, curricular emphasis within any physics program varies over time; as does the researcher's personal categorization.  Therefore that researcher may rate the same data differently if he or she were to re-evaluate the same categorization set again.  Second, the experiment will not be repeatable from one group to another using these methods because each individual experimenter's ideal categorization of the same set will be different, possibly creating a large distortion in the analysis. Third, as \citeauthor{Lin10} found, as topics become more complex, an expert will express uncertainty in his or her own choice, sometimes preferring the choice of another to his or her own.  Finally, if one evaluates each categorization subjectively based on the expected deep structure category for each problem, one assumes the deep structure vs. surface features distinction rather letting that be a conclusion of the statistical analysis. We believe that any groupings should emerge from the data itself.  In other words, the properties and patterns of what makes a categorization expert-like should be an output of the experiment. Similar to outcomes from non-parametric data-mining, it may not always be clear what these characteristics {\it mean}, as they are frequently combinations of many features or latent factors.

\subsection{Visualization of the Data}
Finally,  several studies utilized dendograms to interpret their data, e.g.,~\citeauthor{veld90}~\cite{veld90}. While dendograms are intuitive, they are not very stable. \citeauthor{Milligan80}~\cite{Milligan80} investigated a number of clustering algorithms and compared them using Monte-Carlo generated data from a defined, yet synthetic, cluster model which employed random perturbations.  According to \citeauthor{Milligan80}, complete linkage clustering, a type of dendogram analysis, struggles to recover clusters when there are outliers present in the dataset.  Another type of dendogram analysis, single linkage clustering, is highly sensitive to noise in the dataset. It is for these reasons that it is important to pre-process your data before putting in a subset of your sorters so that you get a dendogram that is clear and interpretable.  However, interpreting a dendogram is a subjective exercise as each dendogram will have a unique threshold where the tree has clustered into groups, yet has not begun to coalesce into a single stem on the tree.  Some dendograms do not have any distinguishable groups at all.  We desired to have an experimental method that required no pre-processing, with a reliable and easily interpretable output suitable for further analysis.  As a result, we have chosen a different approach, based on graphs.  

\subsection{An Alternative Approach}
Given the above concerns, we explored a different model of analyzing and interpreting card-sorting data. To describe clustering on an individual problem level, we decided to approach the analysis as a network. Instead of looking at piles, we decided to look at individual question cards (nodes in the network) and relationships (edges, in this case due to nodes ``being in the same pile''). Networks are well described by graph theory. As the relationship ``being in the same pile'' has no direction (if problem A is in the same pile as B, then B is in the same pile as A), we are looking at undirected graphs. The resulting graphs have the advantage of converting an abstract network into an object that can both be visualized and analyzed using an established canon of mathematical methods.  

As scientists, we prefer simple explanations to complex ones, and sought to distinguish experts from novices using the simplest test possible.  It is for this reason that we compare these categorizations' macroscopic features before continuing on to microscopic features.  The key distinction between the macroscopic and microscopic scales is that the macroscopic scale should not be sensitive to the identity of the problems, while the microscopic scale should be highly sensitive to problem identity.  In choosing mathematical methods for further analysis, we were unexpectedly limited by one feature of  \citeauthor{Chi81}'s and subsequent studies: the ``multiple categorization,'' i.e., the fact that one and the same question card is allowed to be in more than one pile.  This presented a challenge to several existing algorithms. The key measurement we make is a ``distance'' measurement between each pair of categorizations.  Given these distances, we used Principal Components Analysis (PCA) to visualize the data in a few simple plots.  

\section{Visual and Macroscopic Properties of Sample Experimental Data}
We designed and carried out a card-sorting experiment on physics experts and novices at Michigan State University.  A total of 18 physics professors and 23 novices participated in our study.  All of the novices have completed at least the first semester of an introductory physics course at MSU.  We gave each sorter a set of 50 problems to sort based on similarity of solution.  The physics faculty were given the set and allowed to choose a time when they would complete the task at their convenience while the novices were asked to complete the task during a window of a few hours in an informally supervised setting.  Each sorter categorized his or her problems and recorded his groups and group names in a separate packet. Multiple categorization was allowed, but it was in no way communicated to the individual sorters that this practice was expected or endorsed.

%This sample set allowed us to examine human sorting behavior and explore the capabilities of visualization and statistical methods. However, any particular set of data is insufficient in both size and scope to develop general methods, which is our eventual goal. It is common procedure to develop general methods based on artificial models that allow to generate any number of data sets with desired mainstream and outlier properties. We thus use the data and results from this section as the base for developing such a model in Section~\ref{sec:model}.  

\subsection{Visualizing Categorizations as Graphs}\label{subsec:visualize}
\begin{figure}[!ht]
	\includegraphics[height=0.45\textwidth, angle=-90]{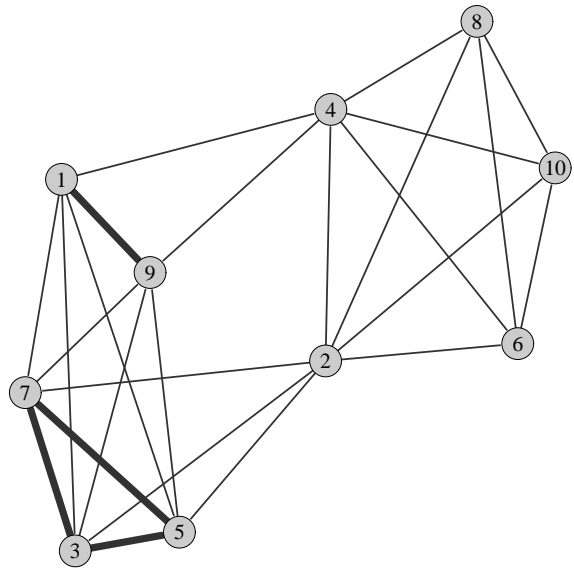}
	\caption{\textbf{Simple example graph:} When two problems are in the same category more than once (problems 1 and 9 as well as problems 3, 5, and 7 in this example) the edges drawn between those two corresponding vertices are thicker.  The line width of each edge was taken proportional to the square of the number of connections between two vertices.  
	\label{ex_graph}}
\end{figure}

Analyzing the experimental data in terms of graphs requires a shift in conceptualization. As a simple example, consider ten questions categorized into four categories.  Suppose that the first category is Newton's second law and contains problems $\{2,4,6,8,10\}$.  Suppose also that the second category is conservation of energy and contains problems $\{1,3,5,7,9\}$, the third category is conservation of momentum and contains problems $\{2,3,5,7\}$, and the fourth category is kinematics and contains problems $\{1,4,9\}$.  At this stage in the process, the names of the categories are irrelevant. In order to create a graph of categorization data we represented question (cards) as the nodes and used each category to create a set of edges.  To start out, we summarize the categorization information in a matrix $T$.  This matrix is a Boolean ($0\left|1\right.$) table with the items being sorted placed along the rows and the categories in each column.  For this example categorization the $T$ matrix is:
\begin{equation*}
T = \left( \begin{array}{cccc}
0		&1		&0		&1\\
1		&0		&1		&0\\
0		&1		&1		&0\\
1		&0		&0		&1\\
0		&1		&1		&0\\
1		&0		&0		&0\\
0		&1		&1		&0\\
1		&0		&0		&0\\
0		&1		&0		&1\\
1		&0		&0		&0  \end{array} \right)
\end{equation*}
This is then converted into a weighted adjacency matrix $X_{ij}$ representing the number of times that item $i$ and item $j$ are in the same category.  Specifically,
\begin{equation}
	X_{ij} = \sum_k T_{ik} T_{jk} \left( 1-\delta_{ij} \right)
\end{equation}
where $\delta_{ij}$ is the Kronecker delta.  Note that $X_{ii}=0$ because in the context of graph theory a term on the diagonal will draw an edge from an object to itself.  Thus, $X_{ij}$ represents the number of edges that must be drawn between two vertices $i$ and $j$ on the graph.  The graph of this example is shown in Figure \ref{ex_graph}. 

Also from the weighted adjacency matrix the adjacency matrix $A_{ij}$ can be derived:
\begin{equation}
	A_{ij} = \min\left(X_{ij}, 1\right)
\end{equation}
We applied this method to the physics problem categorizations created by each sorter.  In doing so, we obtained i) graphs that we may inspect visually ii) adjacency matrices which will be useful for the calculation of certain statistics and iii) weighted adjacency matrices which will be useful when we consider our distance metric.

In order to visualize the graphs seen in Figure~\ref{ex_graph} as well as the other categorization graphs throughout this paper, we utilized the R statistical software's~\cite{Rmanual} igraph package\cite{igraph}.  There are currently 13
different algorithms programmed into R for determining node placement, and
each would cause the same graph to look very differently.  We initially used the Kamada-Kawai algorithm~\cite{kkalgo}, however, we eventually chose the Fruchterman-Reingold  algorithm~\cite{fralgo} because it does the best job of illustrating multiple categorization.  

\begin{figure*}
\begin{tabular}{|c|c|c|c|}
\hline
Sorter 2&Sorter 16&Sorter 20&Sorter 30\\
	\includegraphics[width=0.24\textwidth, angle=-90]{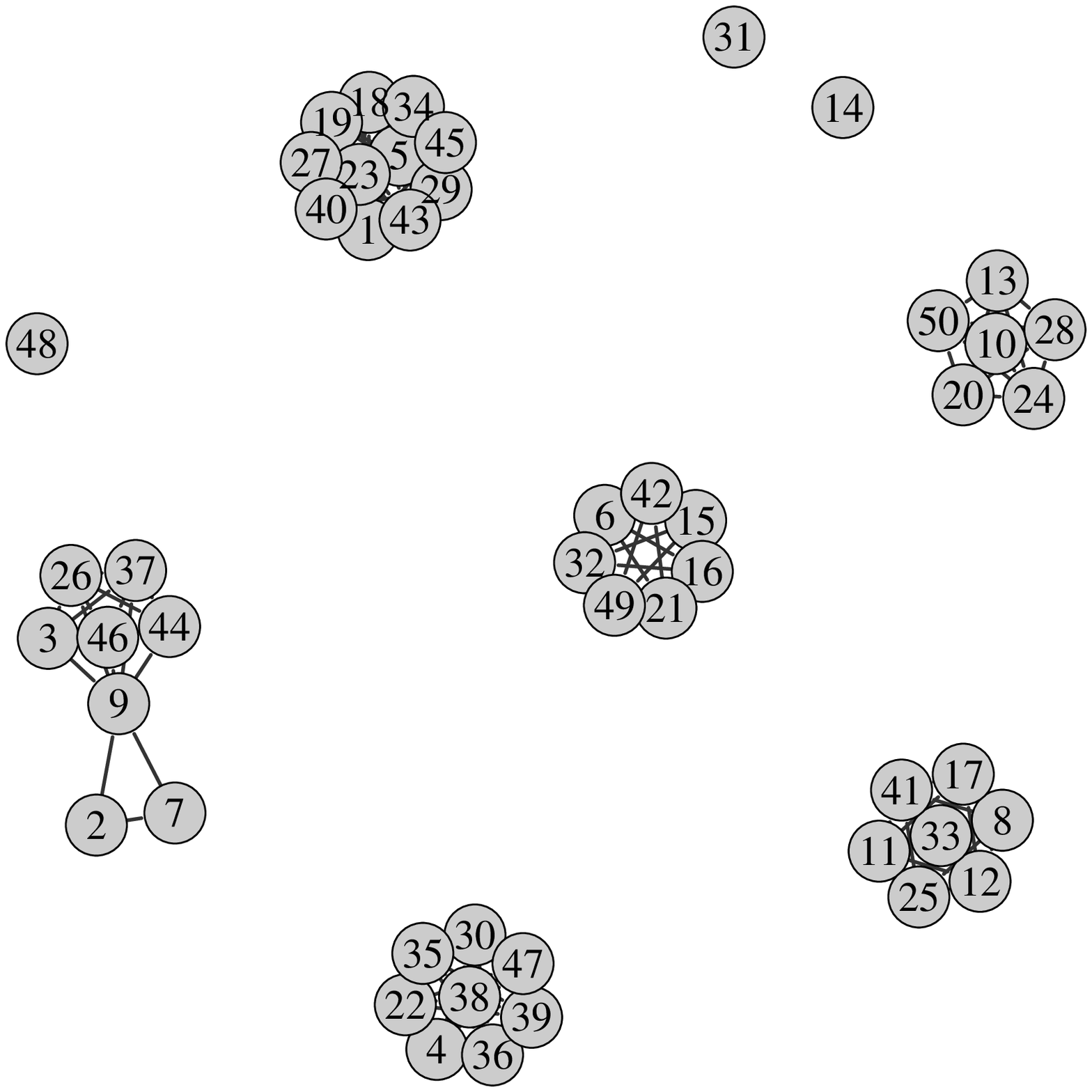}&
	\includegraphics[width=0.24\textwidth, angle=-90]{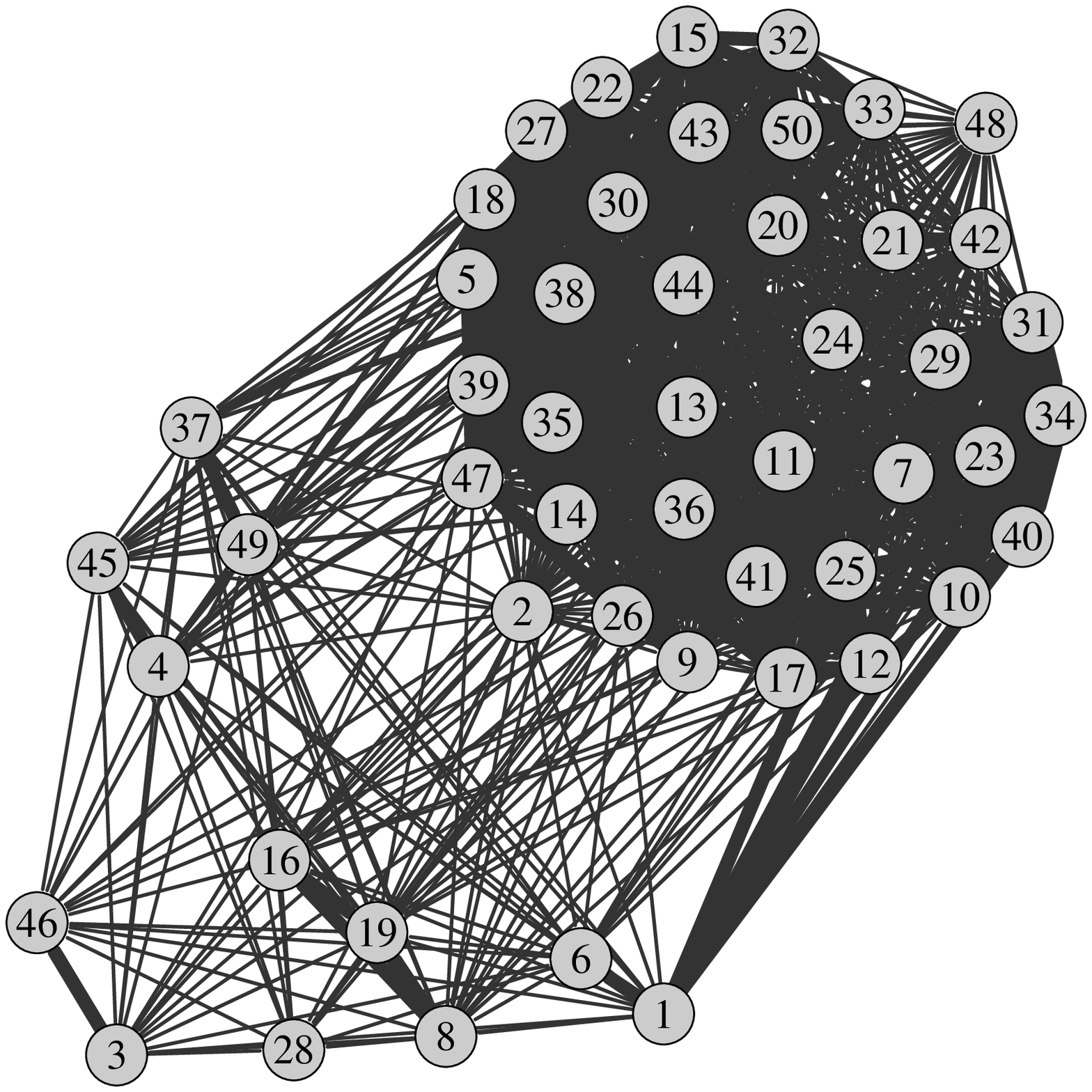}&
	\includegraphics[width=0.24\textwidth, angle=-90]{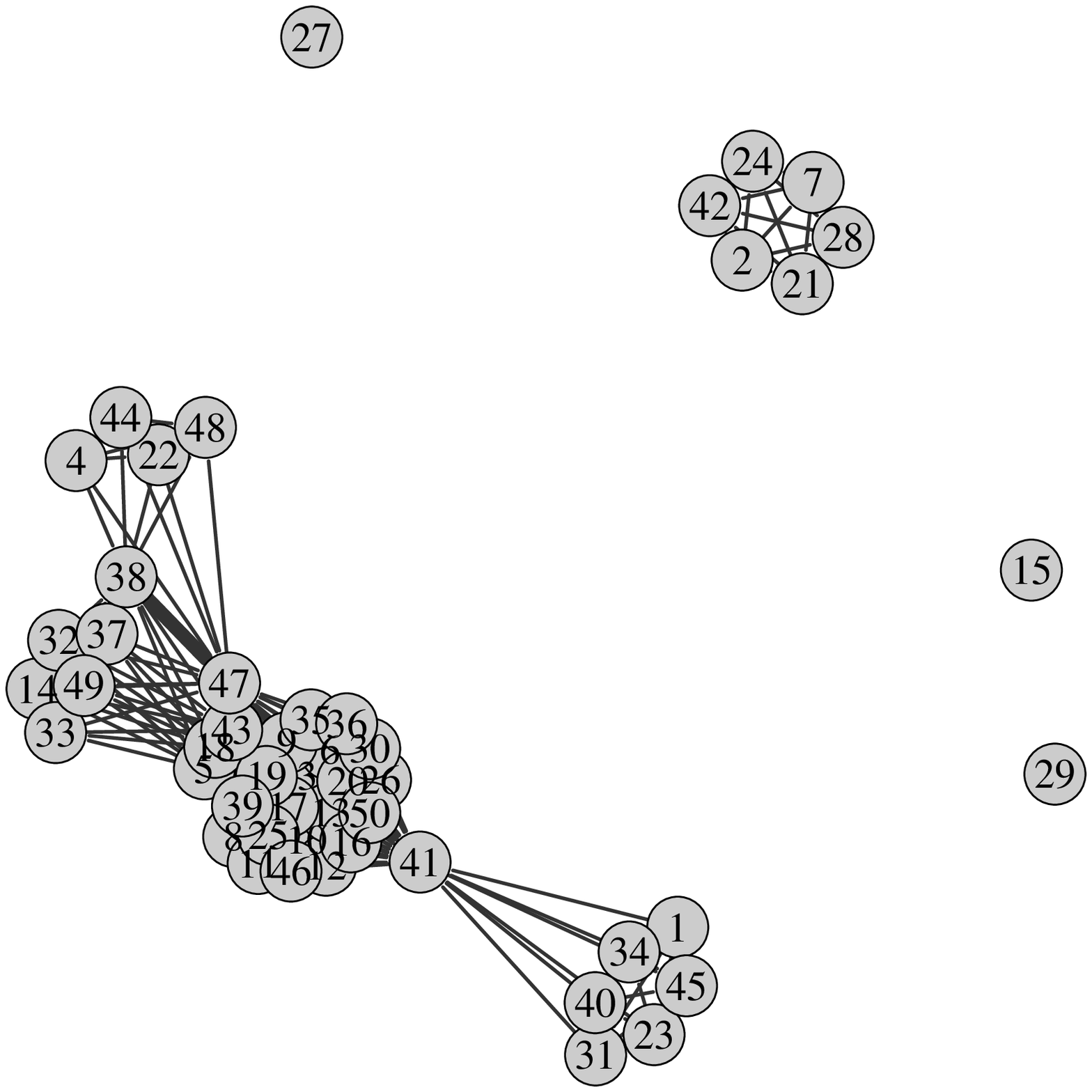}&
	\includegraphics[width=0.24\textwidth, angle=-90]{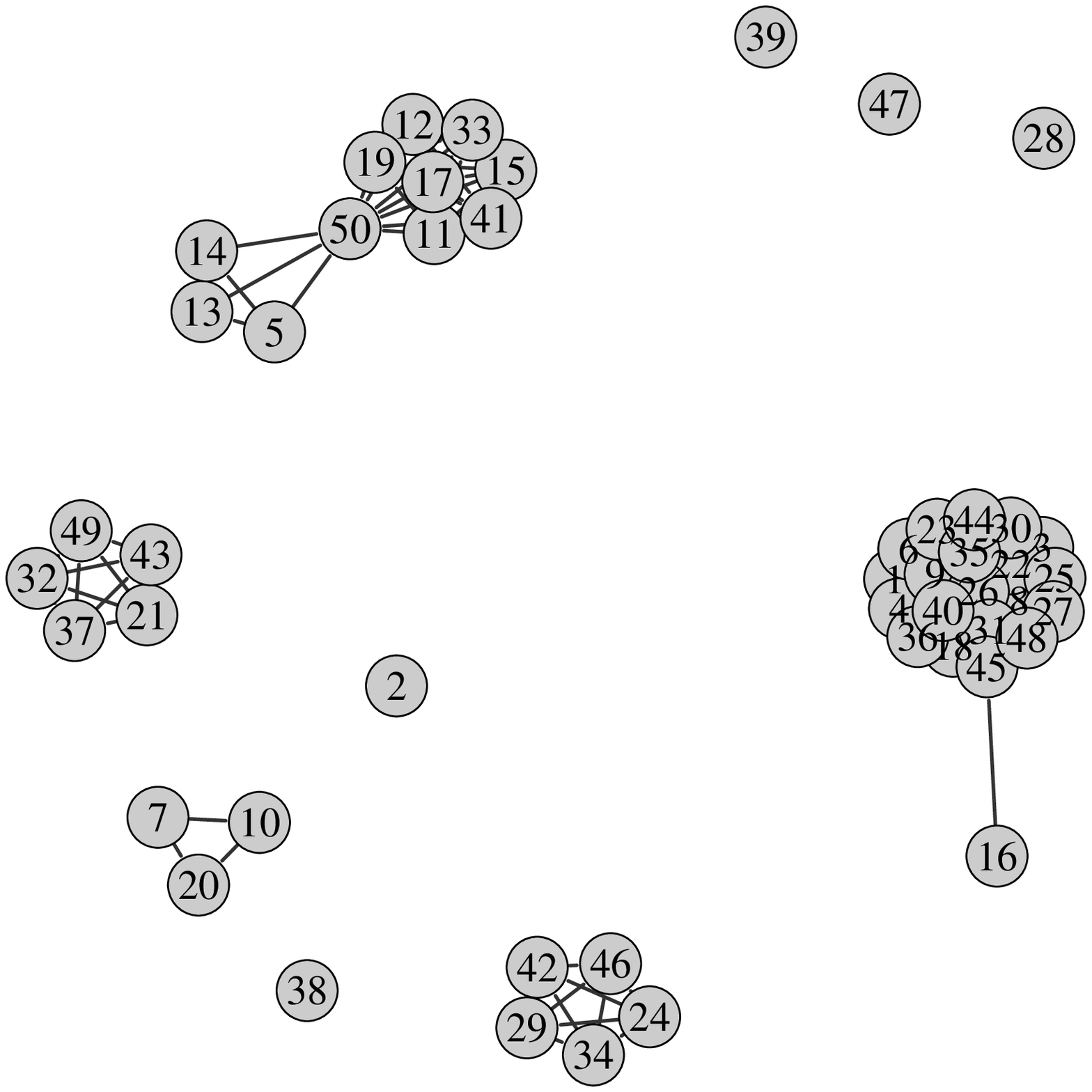}\\
	\hline
	\end{tabular}
	\caption{\textbf{MSU physics study sorter graphs:}
	Displayed from left to right are the categorization graphs for
representative sorters.  Sorters 2 and 16 were
experts and sorters 20 and 30 were novices.  Sorters 2
and 30 did very little multiple categorization, sorter 20 did a good bit of
multiple categorization. Sorter 16 was unique in choosing to categorize each problem between
2 and 3 times.
	\label{sgraphs}}
\end{figure*}

Fig.~\ref{sgraphs} shows the power of the visualization technique: while our sample data had more than 40 participants sorting 50 cards each into any number of piles, flipping through the graphs in less than a minute allowed us to identify the outliers  (such as Sorter 16 in the figure) and general features along which to distinguish the sorters.

\subsection{Number of categories}\label{subsec:numcat}
\begin{figure}
	\includegraphics[height=8cm, angle=-90]{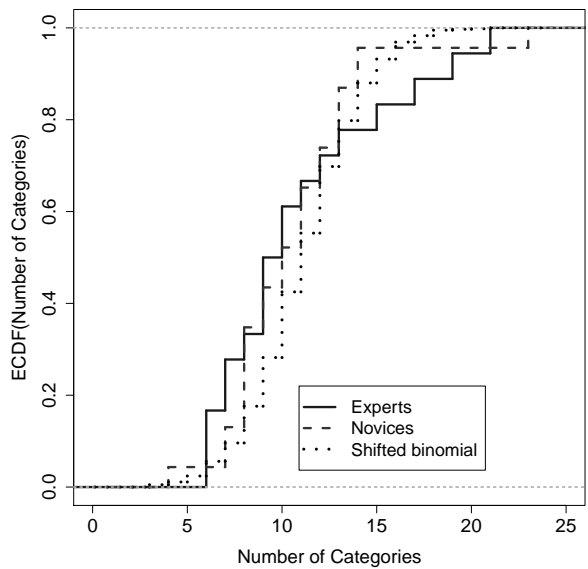}
	\caption{\textbf{Distribution of Number of Categories:} Here we see the ECDFs of the number of category distributions for experts and novices separately.  The faculty set is displayed using the dashed curve and the novice set is displayed using the dotted curve.  We also compare these distributions to a sample $(N=1000)$ shifted binomial distribution with probability $\rho = 0.204$.  A Kolmogorov--Smirnov test comparing these two distributions suggests that expert and novice categorizations are not distinguishable based on category number $(p=0.4793)$, and both are well approximated by the same shifted binomial distribution.
	\label{cat_compare}}
\end{figure}

The ``number of categories'' is a frequently used macroscopic measure of card sorting distributions and not yet particular to graph theory.  It should be noted that \citeauthor{Chi81}'s experiment found that experts and novices created, on average, the same number of categories.  In order to extend this, we perform a test that compares the entire distributions, which includes differences in skewness or shape.  For example, a Gaussian distribution and a bimodal distribution with the same mean and standard deviation would be discriminated in our tests whereas they would not be discriminated when only comparing averages.  In order to compare two distributions, we consider the Empirical Cumulative Distribution Function (ECDF), which is calculated from each normalized distribution $D(x)$ as follows:
\begin{equation}
	ECDF(x) = \int_{-\infty}^x D\left(x'\right) dx'
\end{equation}
For the category number distribution the $ECDF(x)$ represents the fraction of sorters who have $x$ or less categories.  We used the 2-sample Kolmogorov--Smirnov goodness-of-fit hypothesis test (KS-test).  The KS-test statistic is the maximum difference between two ECDFs.  Sample distributions from the same population have a known KS-test statistic distribution.  This allows for the calculation of a $p$-value much in the same way that a $p$-value is calculated from a T-test.  This $p$-value behaves in the usual way: If $p > 0.05$, then the distributions are not statistically different at a 95\% confidence interval.  A KS-test comparing the ECDFs of expert and novice number of categories (see Figure \ref{cat_compare}) demonstrated no statistically significant difference $\left(p=0.4793\right)$.  This result confirms and expands \citeauthor{Chi81}'s result regarding the average number of categories for experts and novices.  Furthermore, we see that these distributions are consistent with a binomial distribution.

\subsection{Connectedness}
\begin{figure}
	\includegraphics[height=8cm, angle=-90]{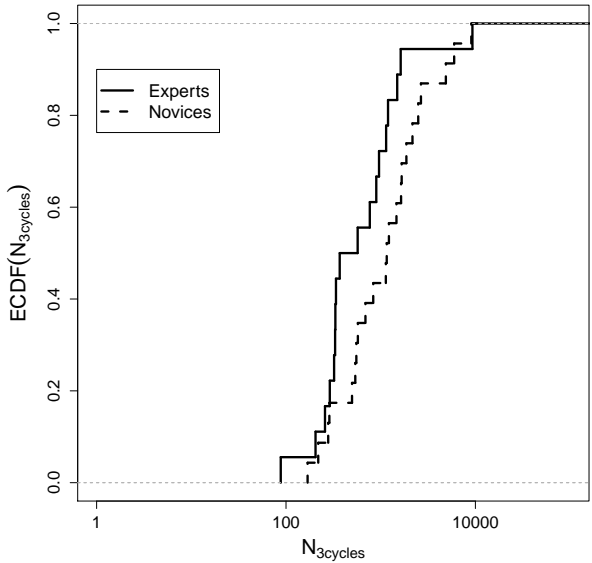}
	\caption{\textbf{Number of 3-cycles:}  This is the distribution of the number of 3-cycles for experts and novices.  A Kolmogorov--Smirnov test suggests that experts and novices are not distinguishable based on their 3-cycle distributions $(p=0.1584)$.  
	\label{en3cyc}}
\end{figure}

The number of so-called 3-cycles macroscopically describes the connectedness of a graph, and is the first graph theoretical measure we apply.  A 3-cycle is a sub-graph of three vertices where all vertices connect by edges.  In our example, shown in Figure \ref{ex_graph}, one of the 24 3-cycles is the sub-graph including vertices $\{1,3,5\}$ because they are all connected by (at least) one edge.  However, the sub-graph including vertices $\{1,2,3\}$ is not a 3-cycle because vertex $1$ is not connected to vertex $2$.  This statistic is related to how often a sorter categorizes cards in multiple piles.  Contrary to the previous example where 7 of the 10 problems were categorized twice, now consider the following example without any multiple categorization.  Suppose the conservation of energy category has problems $\{1,4,7,10\}$, the Newton's Second Law category has problems $\{2,5,8\}$, and the conservation of momentum category has problems $\{3,6,9\}$.  In this categorization, where there are no problems multiply categorized, there are only six 3-cycles.  As such, the 3-cycle distribution is extremely useful for analyzing the connectedness of graphs.  A KS-test comparing the ECDFs of expert and novice 3-cycle distributions (see Figure \ref{en3cyc}) demonstrated no statistically significant difference $(p=0.1584)$.  This result was expected because \textit{connectedness} does not take problem \textit{identity} into account.  

\subsection{Cliques}
\begin{figure}
	\includegraphics[height=8cm, angle=-90]{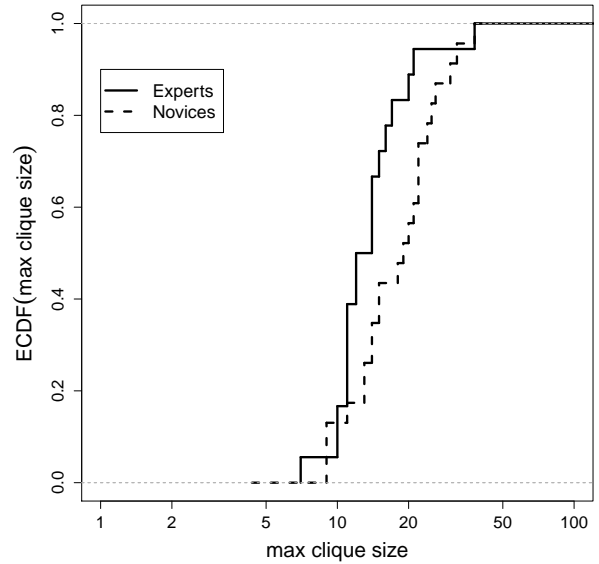}
	\caption{\textbf{Maximum Clique Size:}  This is the distribution of the maximum clique size for experts and novices.  A Kolmogorov--Smirnov test suggests that experts and novices are not distinguishable based on their maximum clique size distributions $(p=0.0587)$.  
	\label{enclique}}
\end{figure}

Our next macroscopic test considers the so-called maximum clique size.  Cliques quantify maximally connected sub-groups.  In our context the maximum clique size is the size of the largest ``pile'' that a sorter has created.  A KS-test comparing the ECDFs of expert and novice maximum clique size distributions (see Figure \ref{enclique}) demonstrated no statistically significant difference  $(p = 0.0587)$. Similar to the \textit{connectedness} result in the preceding section, this result was expected as \textit{maximum clique size} does not take problem \textit{identity} into account.  

\subsection{Diameter}
\begin{figure}
	\includegraphics[height=8cm, angle=-90]{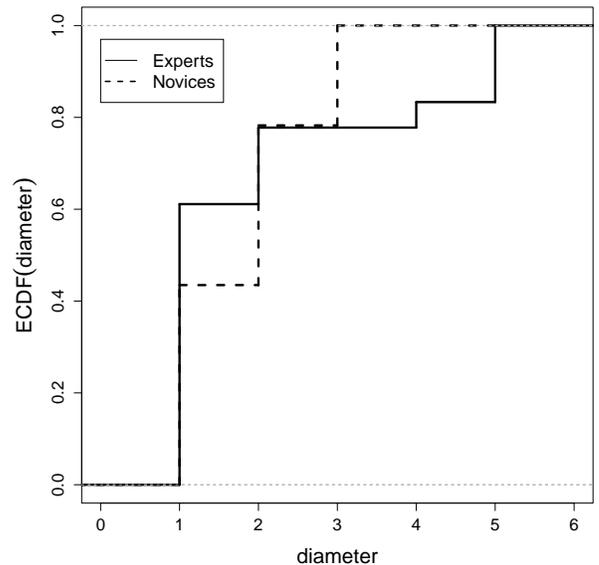}
	\caption{\textbf{Diameter:}  This is the distribution of the diameter of the experts and novices.  A Kolmogorov--Smirnov test suggests that experts and novices are not distinguishable based on their diameter distributions $(p=0.6432)$.  
	\label{endiam}}
\end{figure}

The so-called diameter is a macroscopic measure that describes the number of jumps it takes to get between the two least connected points.  An example of this statistic is the so-called maximum Erd\"{o}s number, which says that many mathematicians can be connected to Paul Erd\"{o}s in 8 steps or less by assuming that two mathematicians are connected if they have collaborated on at least one project.  As such, the diameter distribution is extremely useful for comparing the maximum relative sizes of graphs.  As most of our graphs are unconnected (not every pair of nodes has a path between them), this introduces a difficulty of how to determine the diameter.  While some would choose to find the diameter to be the number of nodes in the graph +1 (or 51 in our case), we chose to ignore all unconnected nodes.  This was done to ensure the largest possible variation in our data.  If we had made the former choice, the ECDF would have (nearly) looked like a step function which would have given the distributions an artificial look, and caused the differences in the data distributions to be almost entirely determined by the outliers, rather than the group as a whole.  A KS-test comparing the ECDFs of expert and novice diameter distributions (see Figure \ref{endiam}) demonstrated no statistically significant difference $(p = 0.6432)$.  This result was also expected as \textit{diameter} does not take problem \textit{identity} into account.  

\subsection{Average Path Length}\label{subsec:avepath}
\begin{figure}
	\includegraphics[height=8cm, angle=-90]{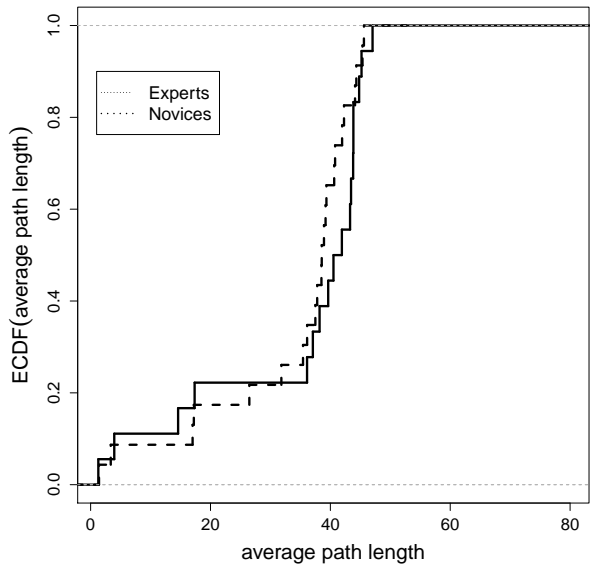}
	\caption{\textbf{Average Path Length:}  This is the distribution of the average path length for experts and novices.  A Kolmogorov--Smirnov test suggests that experts and novices are not distinguishable based on their average path length distributions $(p = 0.3906)$.  \label{enapl}}
\end{figure}

The average path length is a macroscopic measure that describes the average number of jumps it takes to get between all unique pairs of points.  As such, the distribution of average path lengths may be used to compare the average relative sizes of the different graphs.  The calculation of the average path length is subject to the same difficulty due to unconnected graphs as is the diameter.  In this case, we chose to set the path length between unconnected nodes to be 51, rather than ignoring them.  In this setting, we feel that this measure includes both the local structure of the graph and a measure of how unconnected the graph is as well.  As a result, we note that the range of average path length is much larger than the diameter.  However, a KS-test comparing the ECDFs of expert and novice average path length distributions (see Figure \ref{enapl}) demonstrated no statistically significant difference $(p = 0.3906)$.  This result, combined with all of the previous results suggests that our hypothesis that expert and novice categorizations can not be distinguished without taking problem \emph{identity} into account has merit.  

\section{Categorization Models}\label{sec:model}
All of the macroscopic statistical measures, that is, measures which dealt with just the groups of cards and not the individual cards and their identities, yielded no significant distinction between expert and novice sorters. For now, visualizing the data was successful in quickly recognizing outliers (subjects who sort differently), but those outliers were not necessarily more prevalent among experts or novices. We now aim to construct a model of the categorization process that has the same macroscopic and visual properties as our sample experimental data. Along the way, we learn more about human behavior during categorization tasks. 

We started out by using two standard models frequently used in graph theory literature. Unfortunately, neither of these two standard models reproduces the data, in spite of the fact that they are generally considered complimentary. We thus created our own model, which generated more realistic model data.

\subsection{Standard \ER and Barabasi Models} 

An \ER model generates a ``uniform'' graph, that is a graph where any two vertices have a certain fixed probability of  being connected \cite{ErdRen59}.  Uniform graphs may be generated as random realizations of a model having two parameters: the number of nodes and the probability that nodes will connect.  Barabasi graphs, a kind of a ``small-world'' graph often used to model social networking connections\cite{Bara99}, is created by adding one node at a time, and connecting this new node with the existing nodes on the graph with a probability related to the number of edges already connected to each node $P \propto N^a + b$. The model for a Barabasi graph has three parameters, the number of nodes in the graph, the probability to connect to a node with no other connections $(b)$, and the power $(a)$ by which the number of edges already connected to a node $(N)$ is raised.  We describe next the statistical comparison and analysis of graphs generated by these models to the graphs generated by our human sorters.  

\begin{figure*}
\begin{tabular}{|c|c|}
\hline
\ER\ & Barabasi\\
	\includegraphics[width=0.24\textwidth, angle=-90]{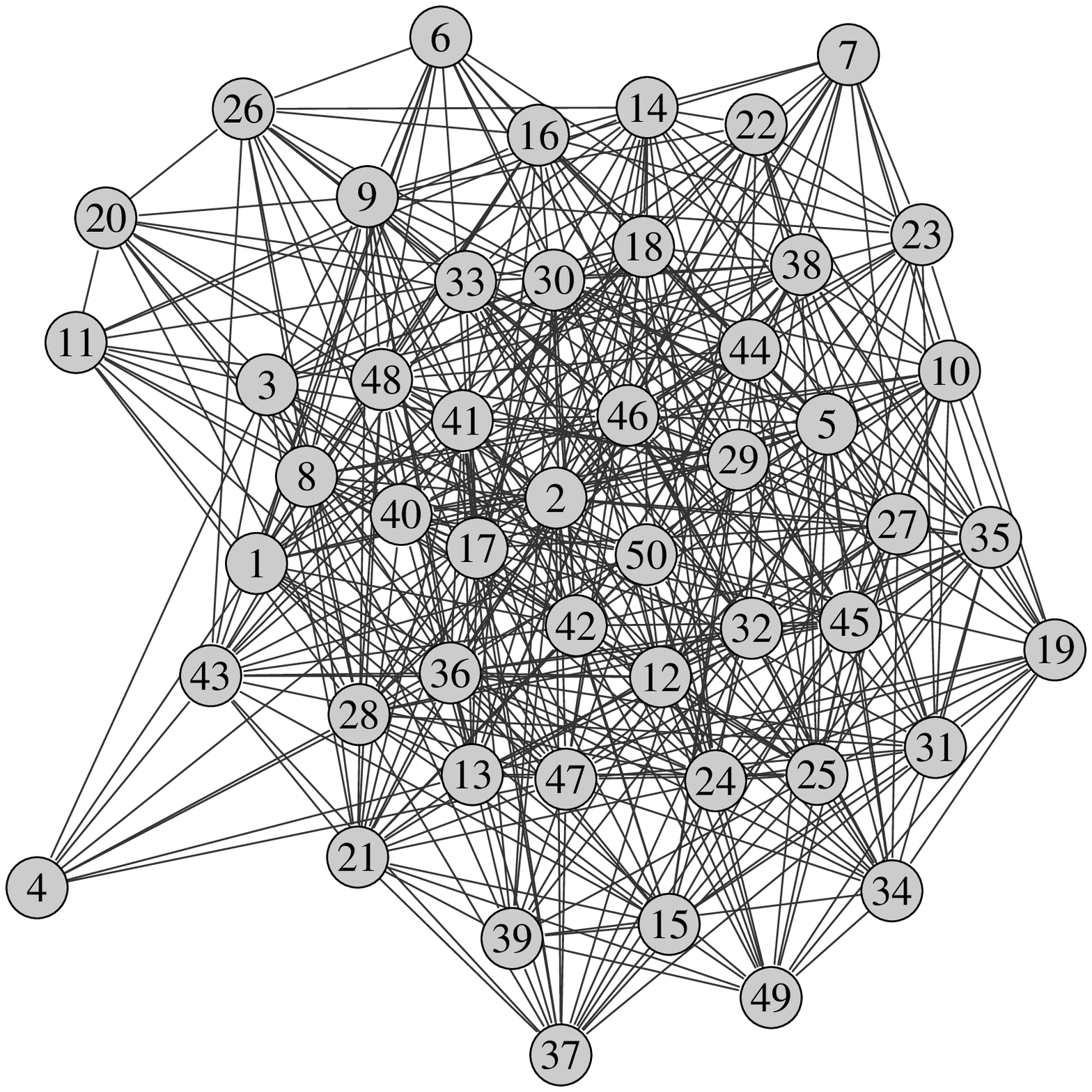}
	\includegraphics[width=0.24\textwidth, angle=-90]{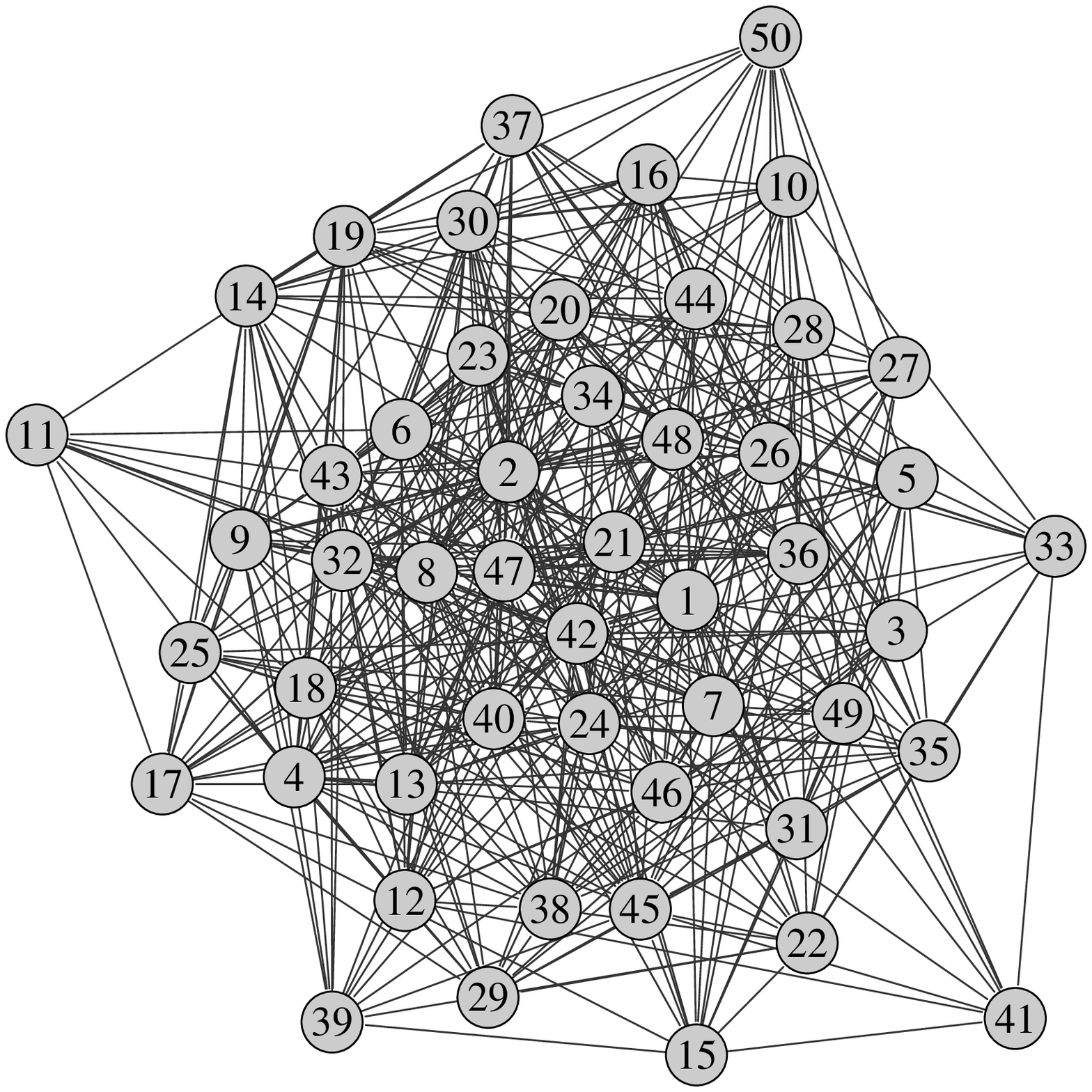}&
	\includegraphics[width=0.24\textwidth, angle=-90]{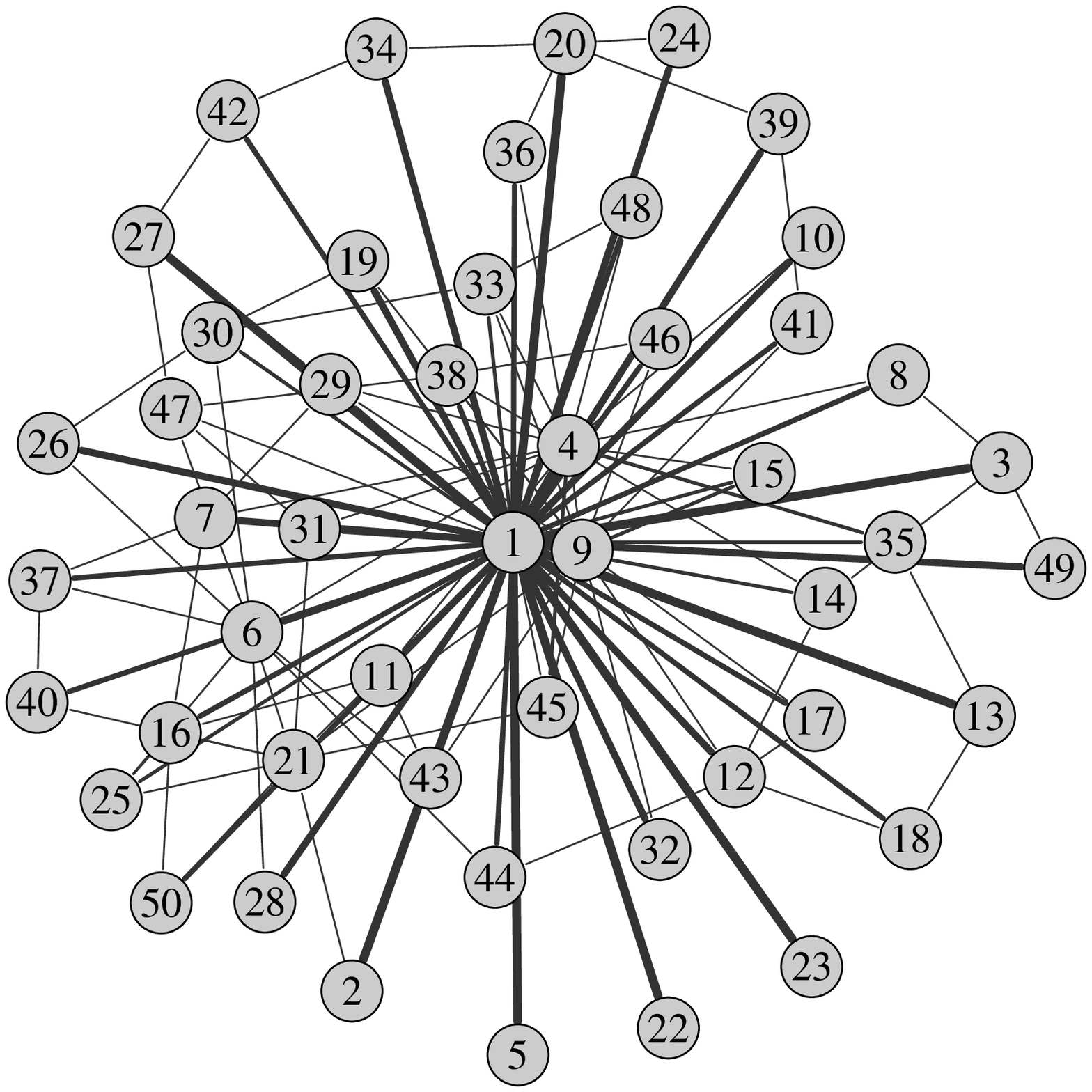}
	\includegraphics[width=0.24\textwidth, angle=-90]{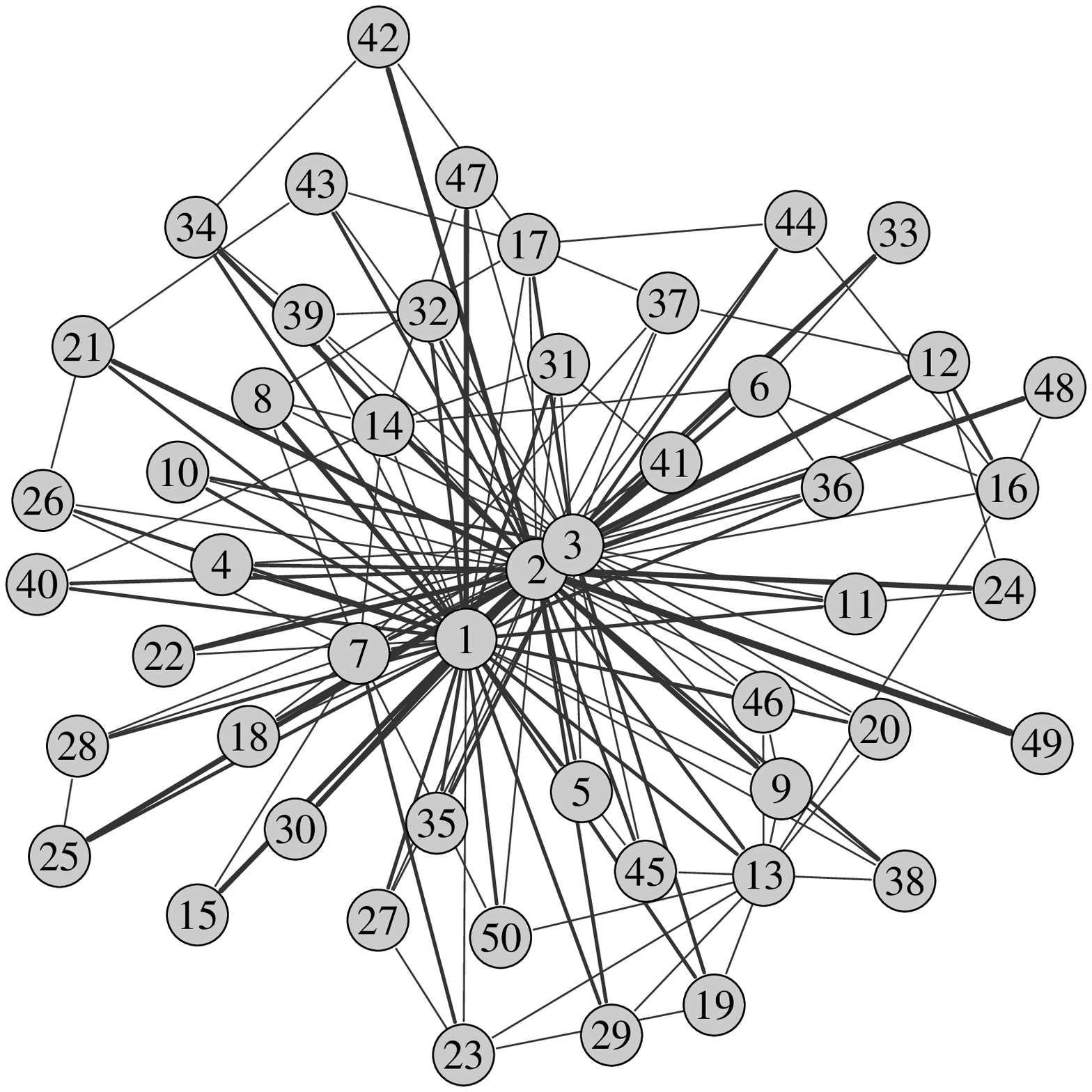}\\
	\hline
	\end{tabular}
	\caption{\textbf{\ER and Barabasi graphs:}  The two graphs on the left are \ER graphs created using optimized parameters that best fit the 3 cycle distributions of experts and novices.  On the right, we see two Barabasi graphs.
	\label{erbagraphs}}
\end{figure*}

\begin{figure*}
	\includegraphics[width=8cm, angle=-90]{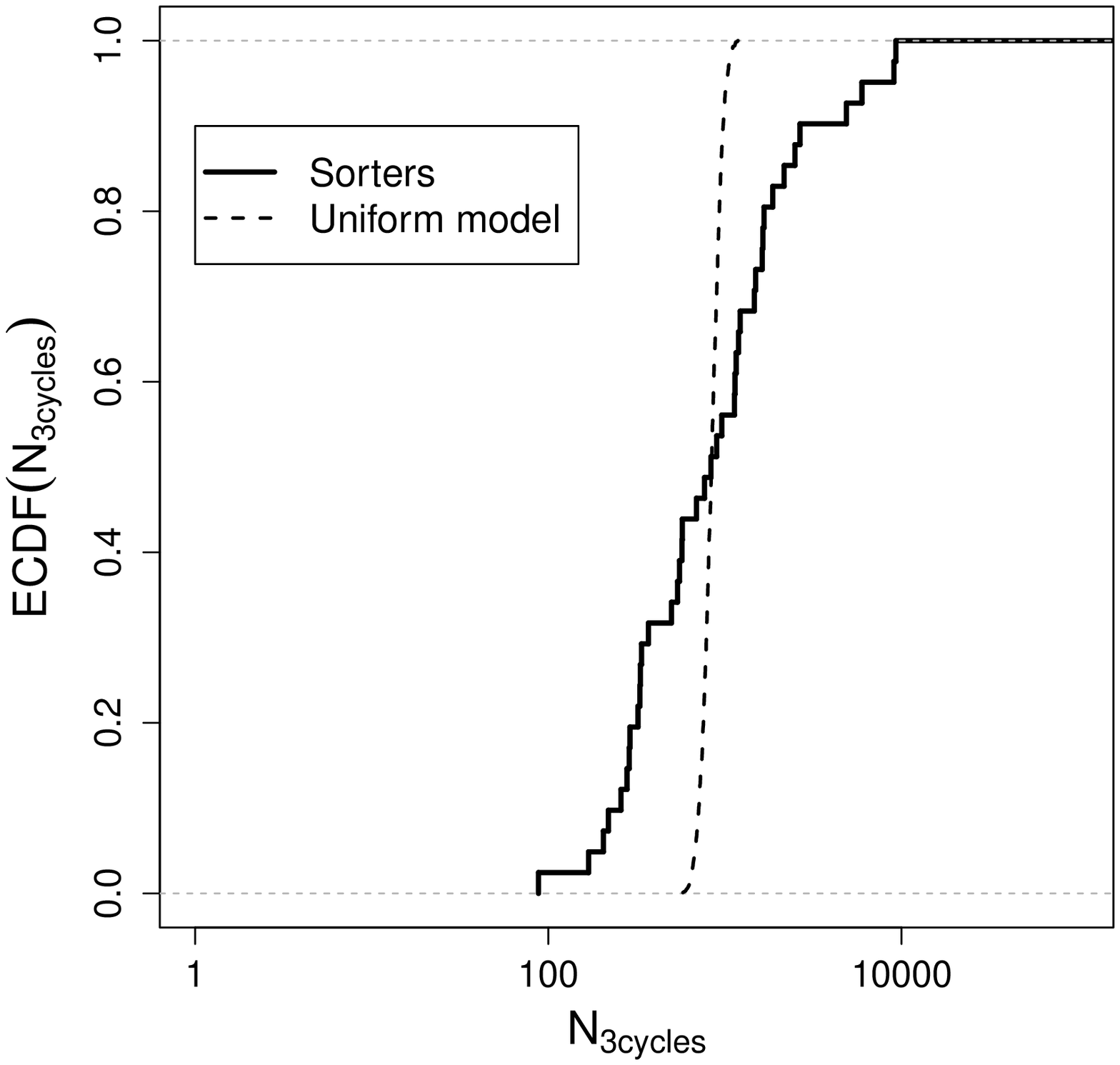}
	\includegraphics[width=8cm, angle=-90]{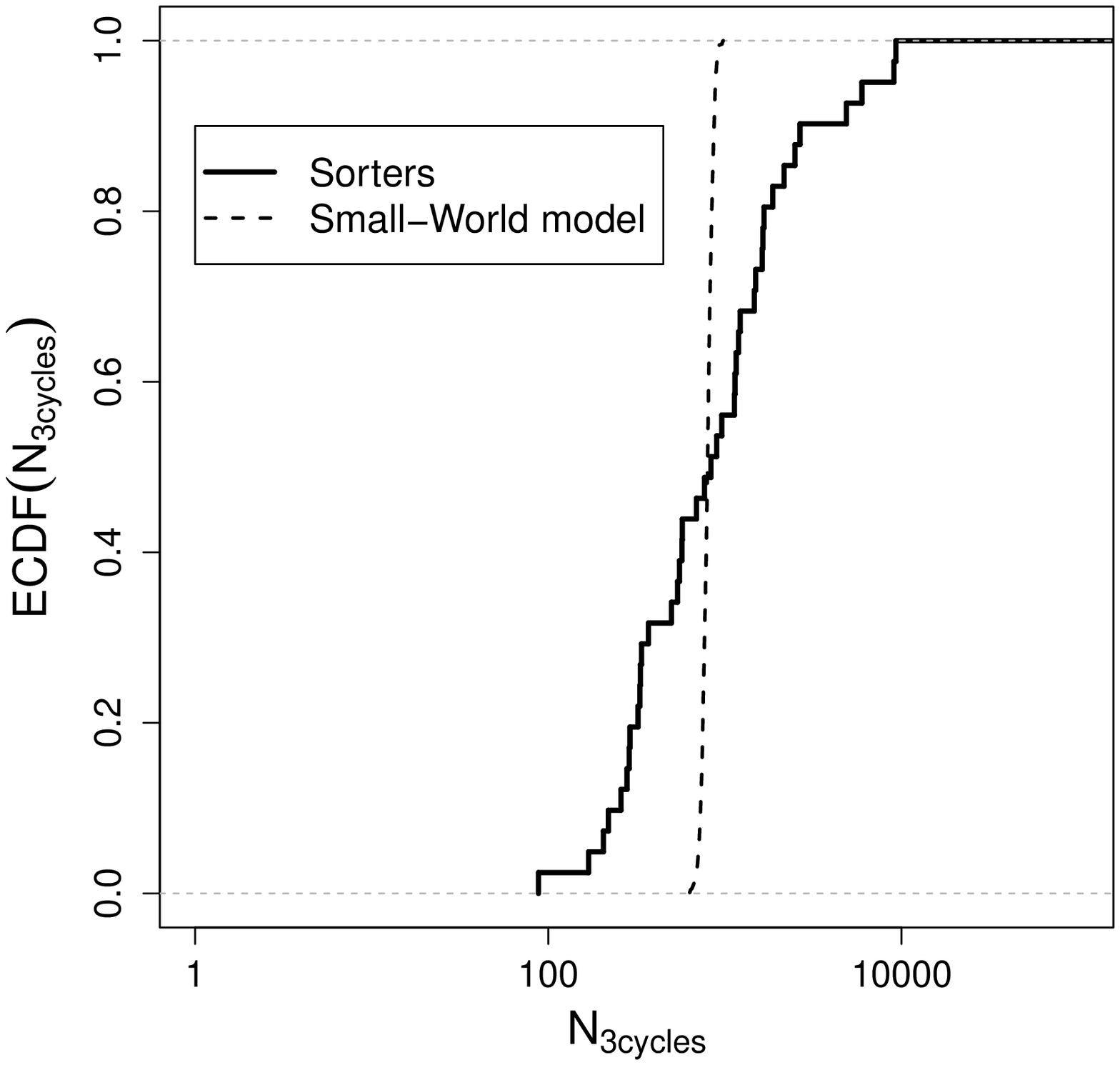}
	\caption{\textbf{3-cycle distributions:}  On the left is an ECDF of the 3-cycle distribution for sorters and the \ER model.  On the right is the corresponding ECDF for the sorters and the Barabasi model.  In both graphs the dashed line corresponds to the appropriate random model while the solid line corresponds to the sorter data.
	\label{gtecdfs}}
\end{figure*}

First, we considered the \ER model.  See Figure \ref{erbagraphs} for examples of \ER graphs.  In order to determine the best input parameters for our model we optimized these parameters using the standard algorithm ``optim'' found in R\cite{Rmanual}.  This was done by calculating 1000 random graphs from the \ER model using test parameters and calculating the 3-cycle distribution from those graphs.  This distribution was then compared to the combined expert and novice 3-cycle distribution from our experiment and we calculated the KS-test statistic for those two distributions.  Ultimately, the parameters that we determined through this optimization for the \ER model were the ones that produced the minimum KS-test statistic between the sorter distribution and the \ER model distribution.  See the left panel of Figure \ref{gtecdfs} for a comparison of the ECDFs for these 3-cycle distributions.  While the optimization was only done for the 3-cycle parameter, the minimum KS-test statistic corresponded to $p<10^{-6}$.  We also compared the sorter distributions to the \ER model distribution for maximum clique size, diameter, and average path length for these optimized parameters.  In every case we found $p < 10^{-6}$ and therefore the \ER model with optimized parameters does not statistically describe the sorter data.  Next we considered the Barabasi model:  See Figure \ref{erbagraphs} for examples of Barabasi graphs.  We repeated the same optimization process for the Barabasi model parameters and also compared the sorter distributions for 3-cycles, maximum clique size, diameter, and average path length.  In every case we find $p < 10^{-6}$ and therefore the Barabasi model does not statistically describe the sorter data.  Due to the difficulty that these canonical models have in describing the sorters' behavior we have chosen to create our own model, which we will call the Cognitive Categorization Model (CCM).  

\subsection{Cognitive Categorization Model (CCM)}\label{subsec:ccm}
As standard models failed to reproduce our experimental data in a satisfactory way, we constructed our own model, which is directly based on the rules of the categorization experiment:
\begin{enumerate}
	\item All questions must be put into a category.
	\item All categories must have at least one question in them.
	\item A question may fall into more than one category.
\end{enumerate}
The latter rule is mathematically cumbersome, but had to be included since it is standard procedure in most experiments, including the one that was the base of our sample data in the previous section.

Our new model, which we call the Cognitive Categorization Model (CCM), has three parameters:
The first parameter of the CCM model $(Q)$ represents the number of questions that are being categorized in the experiment.  The second parameter is the average number of categories determined by a sorter.  As we described in Subsection~\ref{subsec:numcat}, a shifted binomial distribution fits the category number data rather well.  A binomial distribution is the ``weighted coin'' distribution --- if you flip a weighted coin $N$ times, what is the probability that you will get ``heads'' $k$ times?  In principle, one can flip a coin $N$ times and get tails every time.  By Rule \#1, we do not want to allow zero categories, therefore we must introduce a shift.  It would also be senseless to create more categories than questions, so we wish to choose a number of categories from between 1 and $Q$.  The simplest way to do this is to generate a number from the binomial distribution between 0 and $Q-1$ and then add 1 to each of these randomly generated values.  The probability of success is chosen to correspond to the final average number of categories.  The final parameter is the probability to categorize a card into more than one pile.  After each problem has been sorted into a single pile, the algorithm tests whether that problem should be sorted into other categories as well.  Our model, with 2 free parameters is on par with the \ER model (1 free parameter) and the Barabasi model (2 free parameters).  In addition to the fact that the CCM parameters are interpretable, the small number of CCM parameters makes this model parsimonious.  

Appendix~\ref{pcode} shows the pseudocode for this model.  In our code, we implement multiple categorization by generating a random number between zero and one and comparing that number to our multiple categorization probability.  However, there are a number of ways that we can model the multiple categorization probability.  The simplest way is to allow every sorter to have a uniform probability and say that some percentage of the time a card will be split again.  So for this model the multiple categorization probability is constant:
\begin{equation}
	P_{\text{multiple}} = \beta_1
\end{equation}
where $\beta_1$ is a constant between zero and one which applies to the entire population.  Another way that we consider assumes that a penalty is incurred whenever a card is split:
\begin{equation}
	P_{\text{multiple}} = \beta_2^{N}
\end{equation}
where $\beta_2$ is a constant between zero and one which applies to the entire population and $N$ is the number of times that a problem has already been categorized by a random sorter.  Finally, we consider a model where the multiple categorization probability depends on the number of categories $(C)$ that a sorter has selected which was determined by the binomial distribution.
\begin{equation}
	P_{\text{multiple}} = \beta_3^{C}
	\label{ccmv3}
\end{equation}
where $\beta_3$ is a constant between zero and one which applies to the entire population.  The differences between these three choices are so subtle that we cannot see a difference between them by eye using the graphical representation.  

\begin{figure}
	\includegraphics[height=8cm, angle=-90]{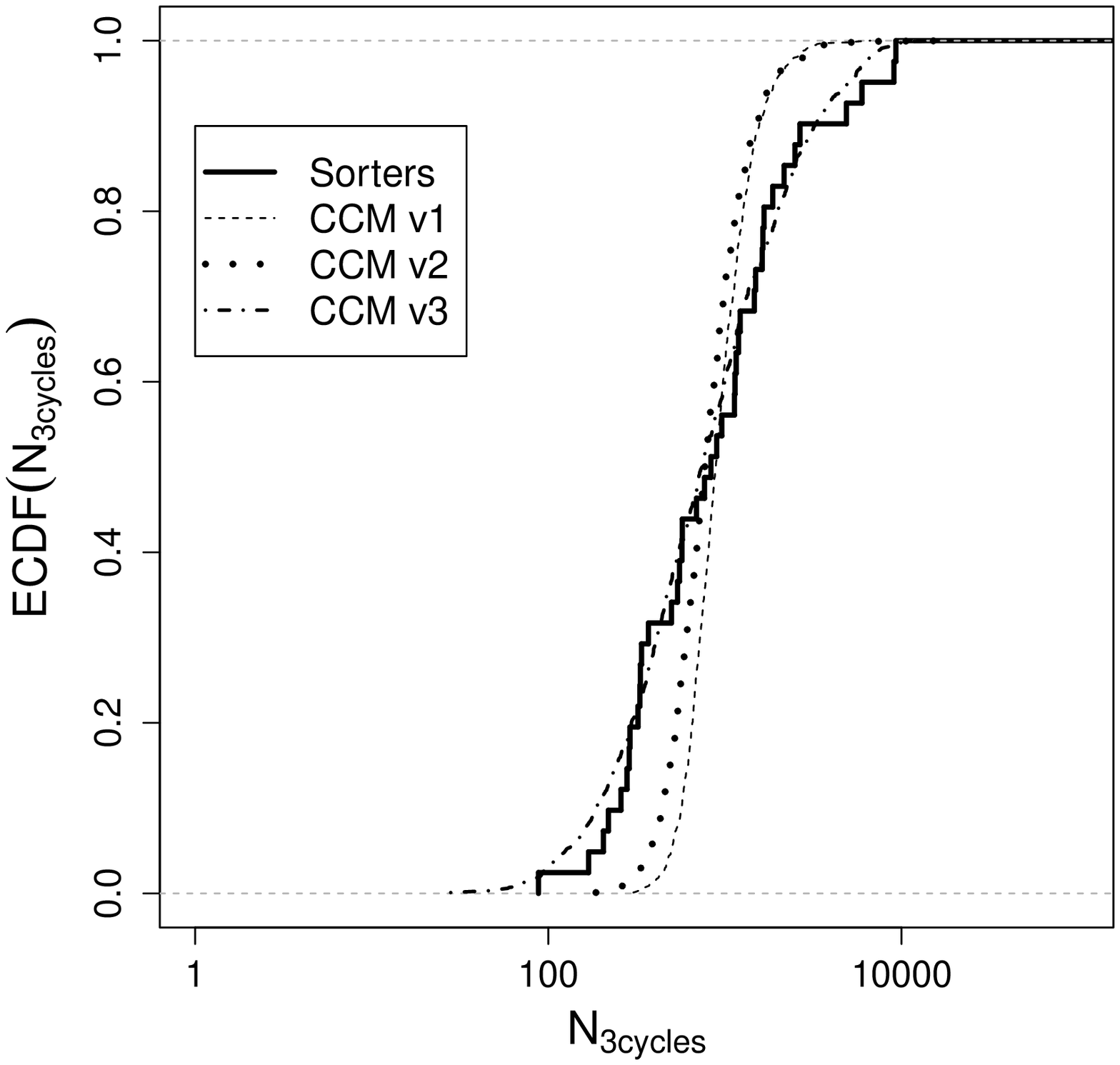}
	\caption{\textbf{3-cycle distributions} Here we see the 3-cycle distributions for the different CCMs.  The model that fits the best is v3, where the multiple categorization probability is $\beta^C$.
	\label{ccmecdfs}}
\end{figure}

In order to determine best-fitting parameters for each of the models we considered, we minimized the KS-test statistic between the data 3-cycle distribution and the model 3-cycle distribution.  For the CCM, we used a simple brute-force grid search instead of the standard optimization algorithm found in the R statistical software\cite{Rmanual}.  The reason for this difference was that the 3-cycle distribution was better approximated with smaller sample sizes for the two standard graph theory models.  However, running the standard optimization algorithm for the larger sample sizes required by the cognitive categorization model took much longer and the brute force method quickly became preferable as we could use smaller sample sizes to get some coarse grained resolution.  Later, we then used larger sample sizes when we got close to the end result.  Once we obtained optimized parameters for the different CCMs, we compare them (see Figure \ref{ccmecdfs}) to the human sorters based on the 3-cycle distribution.

\begin{figure*}
	\includegraphics[width=0.24\textwidth, angle=-90]{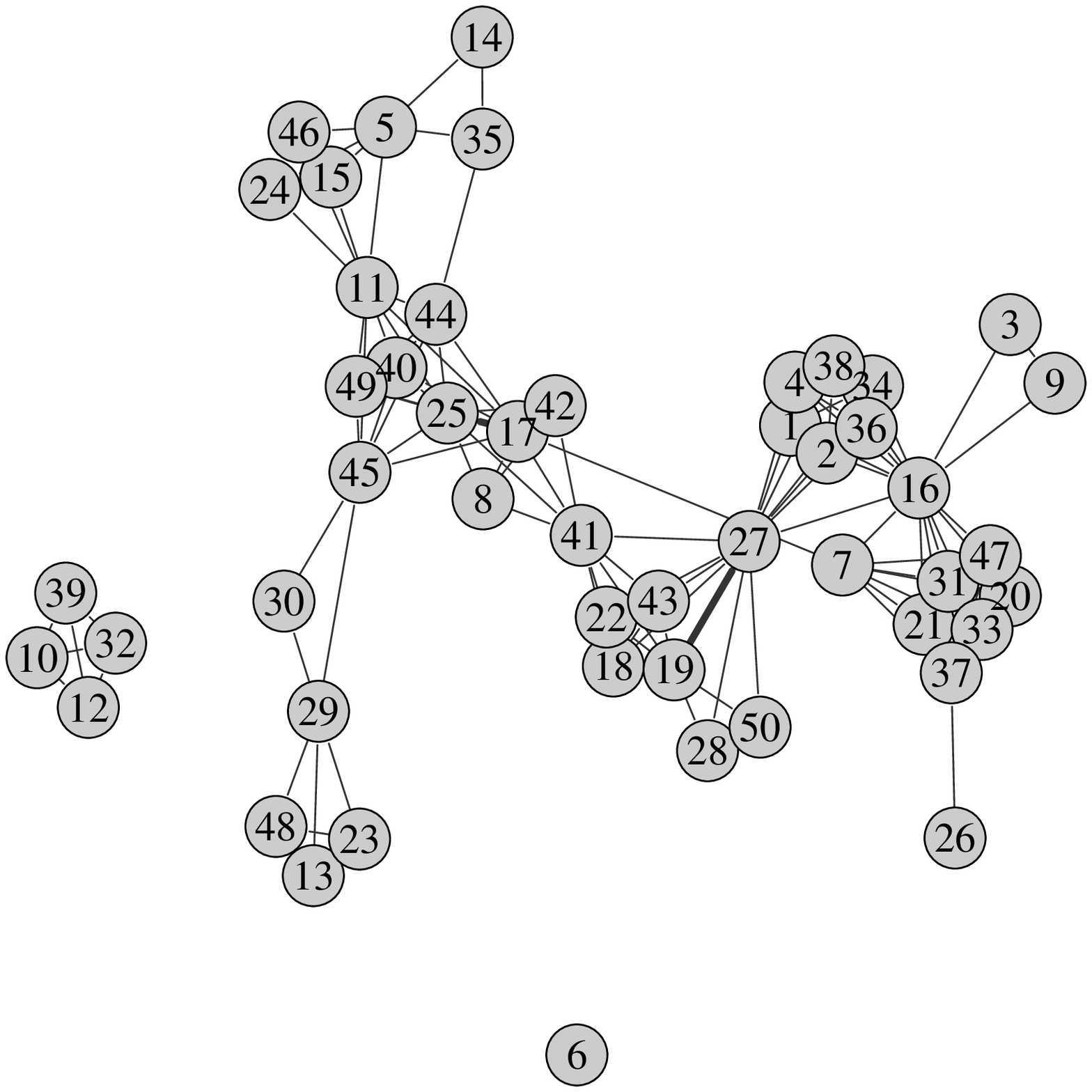}
	\includegraphics[width=0.24\textwidth, angle=-90]{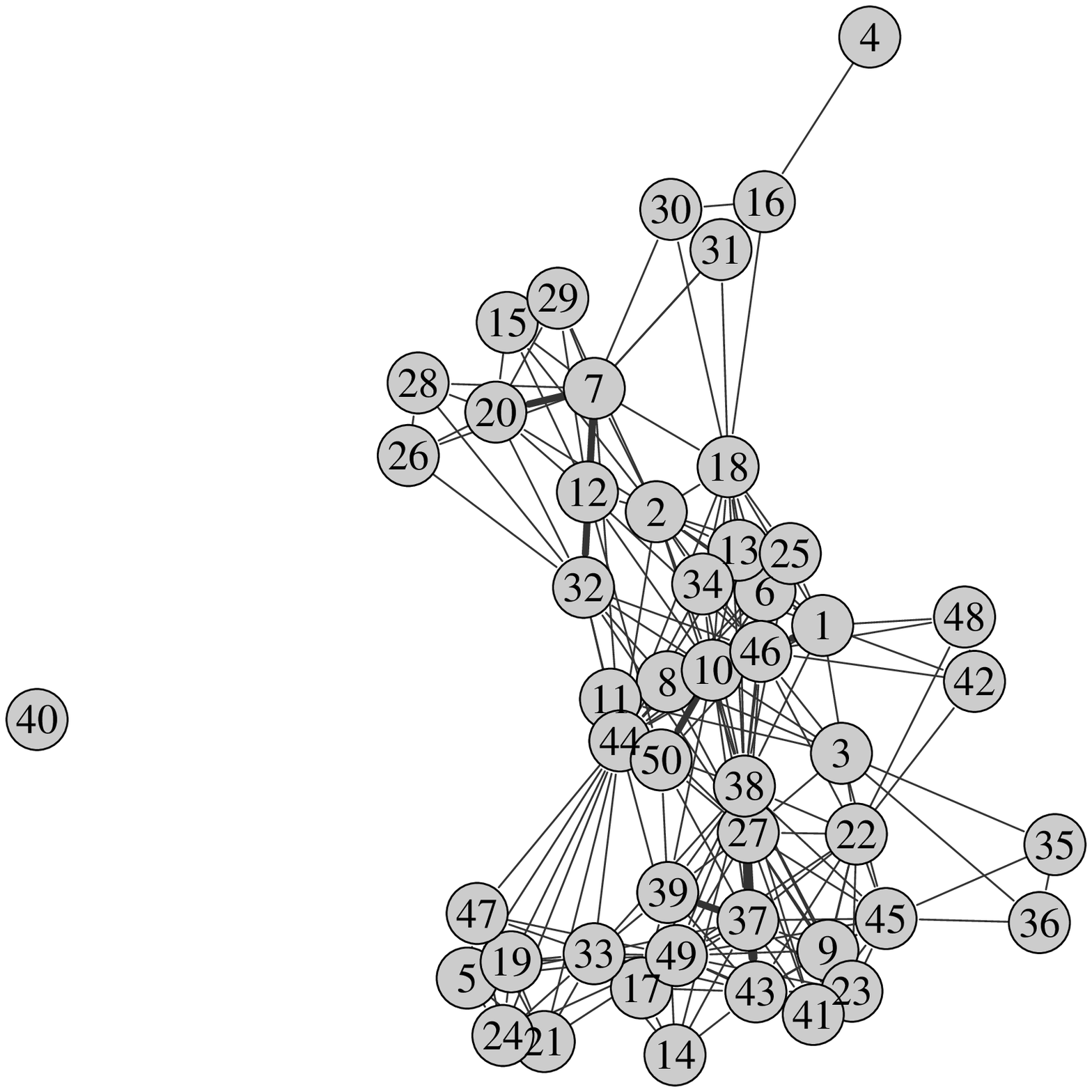}
	\includegraphics[width=0.24\textwidth, angle=-90]{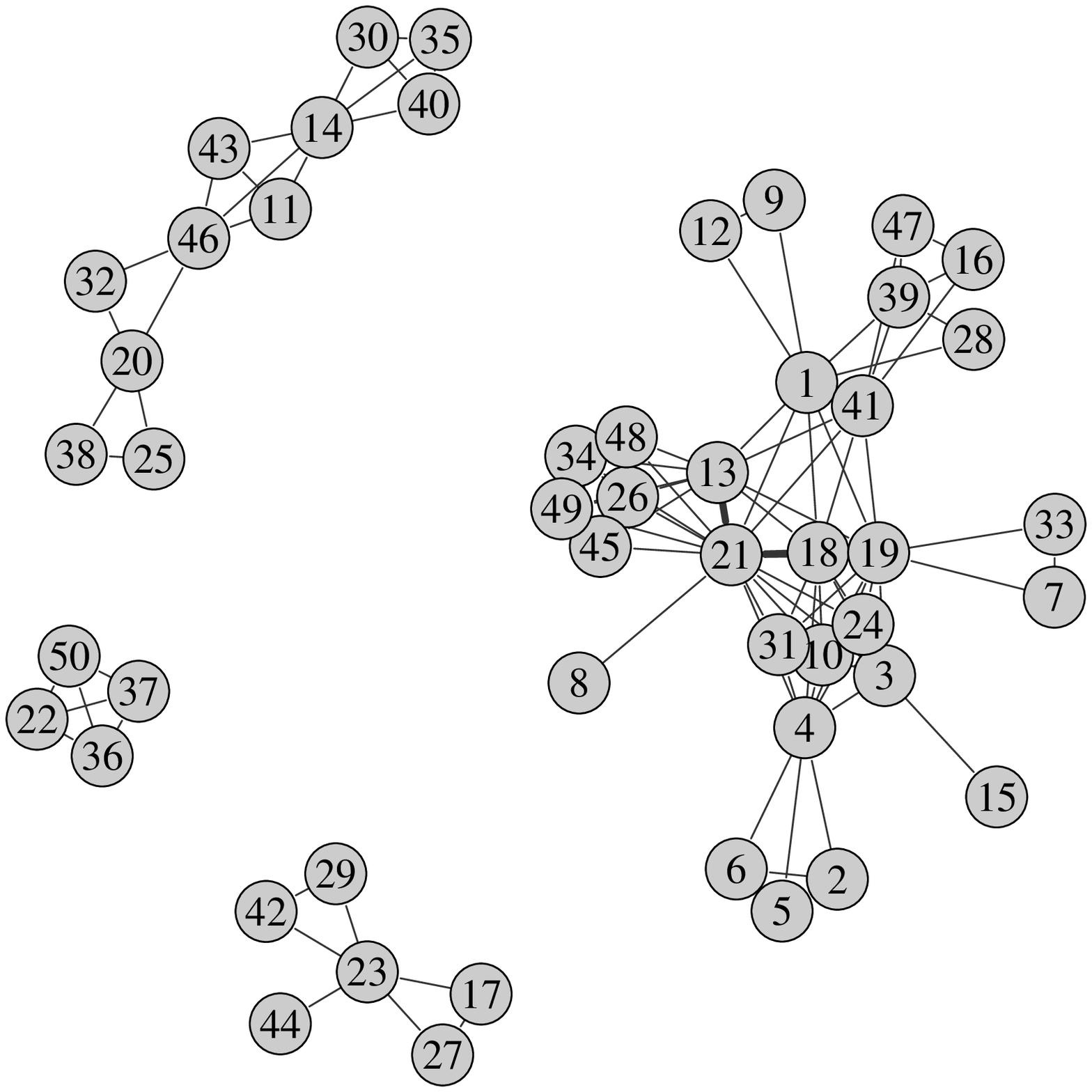}
	\includegraphics[width=0.24\textwidth, angle=-90]{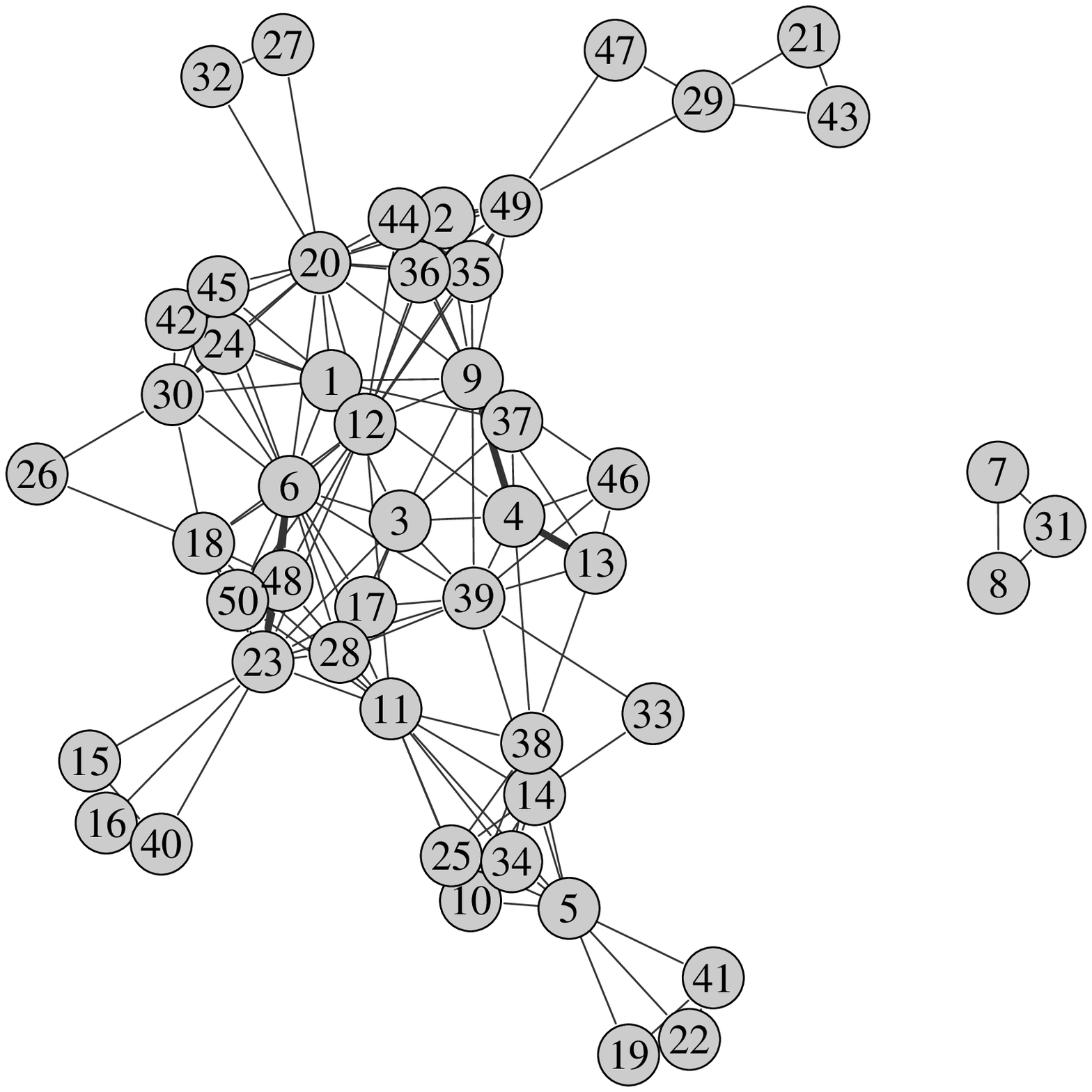}
	\caption{\textbf{Representative CCM graphs:} These are some representative graphs for the CCM using optimized input parameters.  Qualitatively they match up much better to the sorter graphs seen in Figure \ref{sgraphs}.
	\label{ccmgraphs}}
\end{figure*}

We found that the best fitting CCM had a multiple categorization probability that depended on the number of categories as seen in Equation~\ref{ccmv3}.  Figure \ref{ccmgraphs} shows us some example CCM graphs displayed with optimized input parameters for the model that we eventually select.  These CCM graphs shown look much more like the graphs of the human sorters seen in Fig.~\ref{sgraphs}.  

The success of this model gives us insight into the behavior of our sorters as the probability to categorize a single problem in multiple categories is different for each person and that probability actually decreases with the number of categories that are created.  This model prediction is supported by an observation we made of our sorters.  We observed two different sorter behaviors while they worked on the categorization task.  It seemed like some people were resolved to make as few piles as possible, we will call these people ``stackers.''  Stackers were more likely to put a problem in multiple categories, deciding that putting a problem into two piles was a better decision than making a new category.  As a result a stacker's groups tend to be large and inclusive.  The other group of people would spread the problems out on the table that they were working on, we will call these people ``spreaders.''  Spreaders were less likely to put a problem in multiple categories, deciding that making a new category was a better decision than putting a problem into two piles.  As a result a spreader's groups tend to be small and exclusive.

\section{Principal Components of the Distance Metric}
\subsection{Distance Metric}
We now create a distance metric as a microscopic measure to compare two sorters to  each other.  There are several existing distance metrics that will compare two different sortings, including statistical indices such as the Rand Index~\cite{Rand_orig} which can be converted into a distance metric.  However, in searching for existing statistical methods that will work for our categorization exercise, we found none that obeyed the rules of our ``categorization game,'' especially the third rule.  The Rand index merely counts the number of ``agreements.''  This may be calculated for any two categorizations.  However similarity indicies that are not corrected for chance agreements are not as reliable for creating a sort of measuring stick for measuring a ``distance'' between two categorizations \cite{rand2, chance_corr}.  For this reason \citeauthor{rand2} created an adjustment to the Rand index.  However this adjustment requires that the sub-groups are \textit{disjoint}, eliminating any utility that the adjusted Rand index has for our study and other similar studies which allow multiple categorization.  This story may be repeated for any one of the other statistical indicies that we could find in the statistics literature, and after some consideration, we decided that we needed to invent a new method for analyzing this type of data.  

Our distance metric will bypass this difficulty as it is a direct distance metric and not a similarity index.  The distance metric is determined by considering the weighted adjacency matrix for each reviewer, and it compares any two graphs generated by two reviewers as long as they have the
same number of nodes (which they would for identical card sets). Each element of the weighted adjacency matrix $X^r$ for each reviewer $r$ is:
\begin{equation}
	X_{ij}^r = \text{number of edges between problems }i\text{ and } j
\end{equation}
The distance metric is:
\begin{equation}
	d_{rs} = \frac{1}{2} \sum_{i=1}^Q \sum_{j=1}^Q \left| X_{ij}^r - X_{ij}^s \right|
	\label{eq:dmetric}
\end{equation}
Our distance metric may be interpreted as the number of edges that need to be added to and removed from one graph to make it identical to another.  The factor of $\frac{1}{2}$ is included due to the symmetry of the weighted adjacency matrix.  In order to be a statistical distance metric, the distance $d$ between any two categorizations $X^r$ and $X^s$ must satisfy a few properties:
\begin{eqnarray}
	d_{rs} &\geq& 0 \nonumber\\
	d_{rs} &=& 0 \quad \Longleftrightarrow \quad X^r = X^s \nonumber\\
	d_{rs} &=& d_{sr} \nonumber\\
	d_{rt} &\leq& d_{rs} + d_{st}
\end{eqnarray}

Appendix \ref{proof} contains a brief proof that our metric satisfies these properties.  We can use this distance metric to create a symmetric matrix where the distance between sorter $i$ and $j$ appears in row $i$ and column $j$.  

\subsection{Principal Component Analysis}
The distance matrix we constructed answers the question ``How far is sorter $i$ from sorter $j$?" for every pair of sorters. Since we have 41 sorters, this matrix operates in a 41-dimensional space, which is of course impossible to visualize. Principal Component Analysis (PCA) is a way of reducing high dimensional data back down to something
more manageable.  PCA is a general term in the statistical community describing a number of techniques involving the singular value decomposition.  To visualize our data, we did a singular value decomposition on the distance matrix.  By applying the singular value decomposition to the distance
matrix we perform a change of base so that the largest amount of
variation is in the first principal component (PC1), the second largest amount of
variation is in the second principal component (PC2), and each subsequent component
explains less variation than the previous component.  This analysis is then
projected out onto fewer spatial dimensions, using only the most influential base vectors as a new reduced base.

\begin{figure*}[!ht]
	\includegraphics[width=8cm, angle=-90]{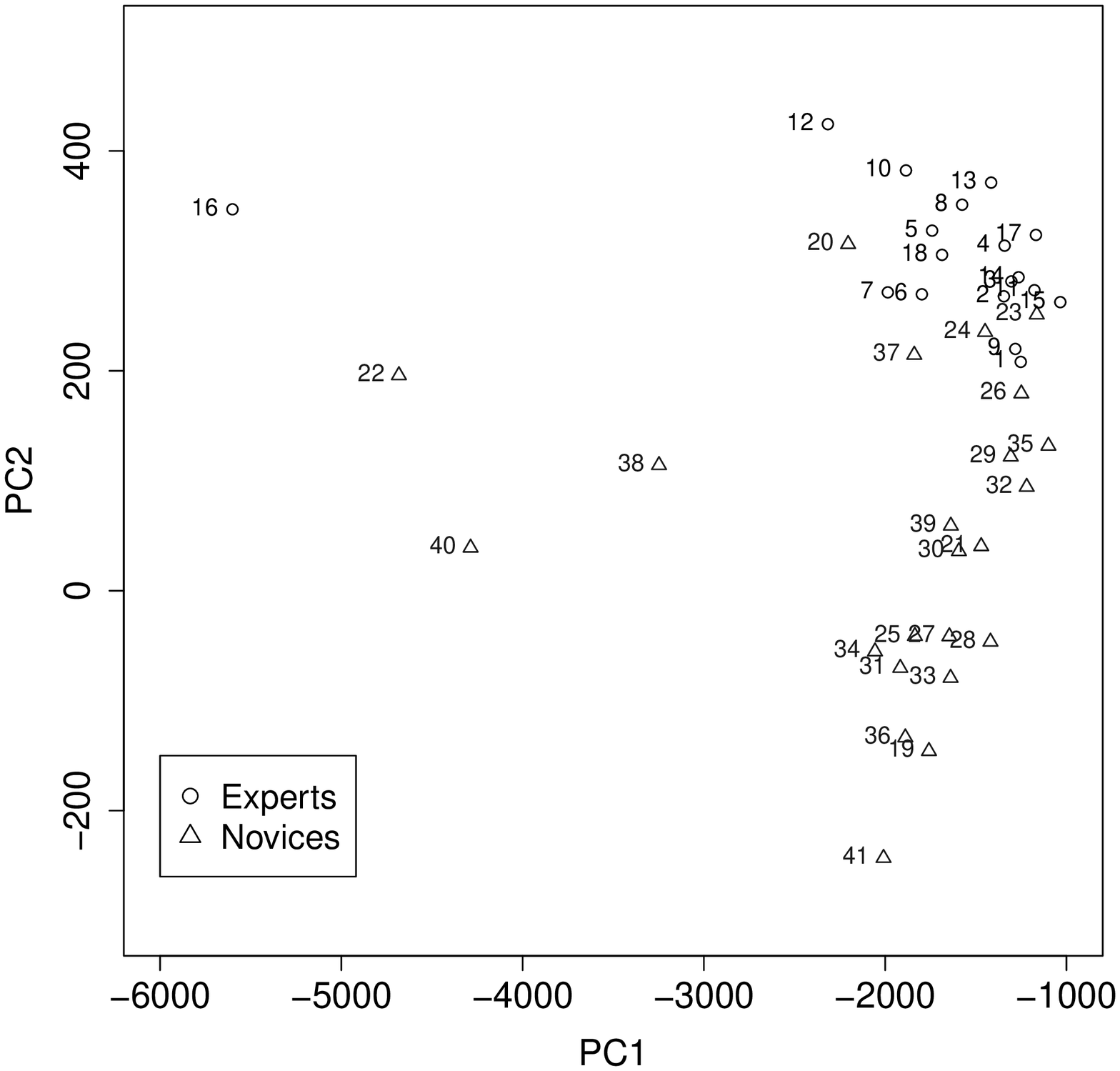}
	\includegraphics[width=8cm, angle=-90]{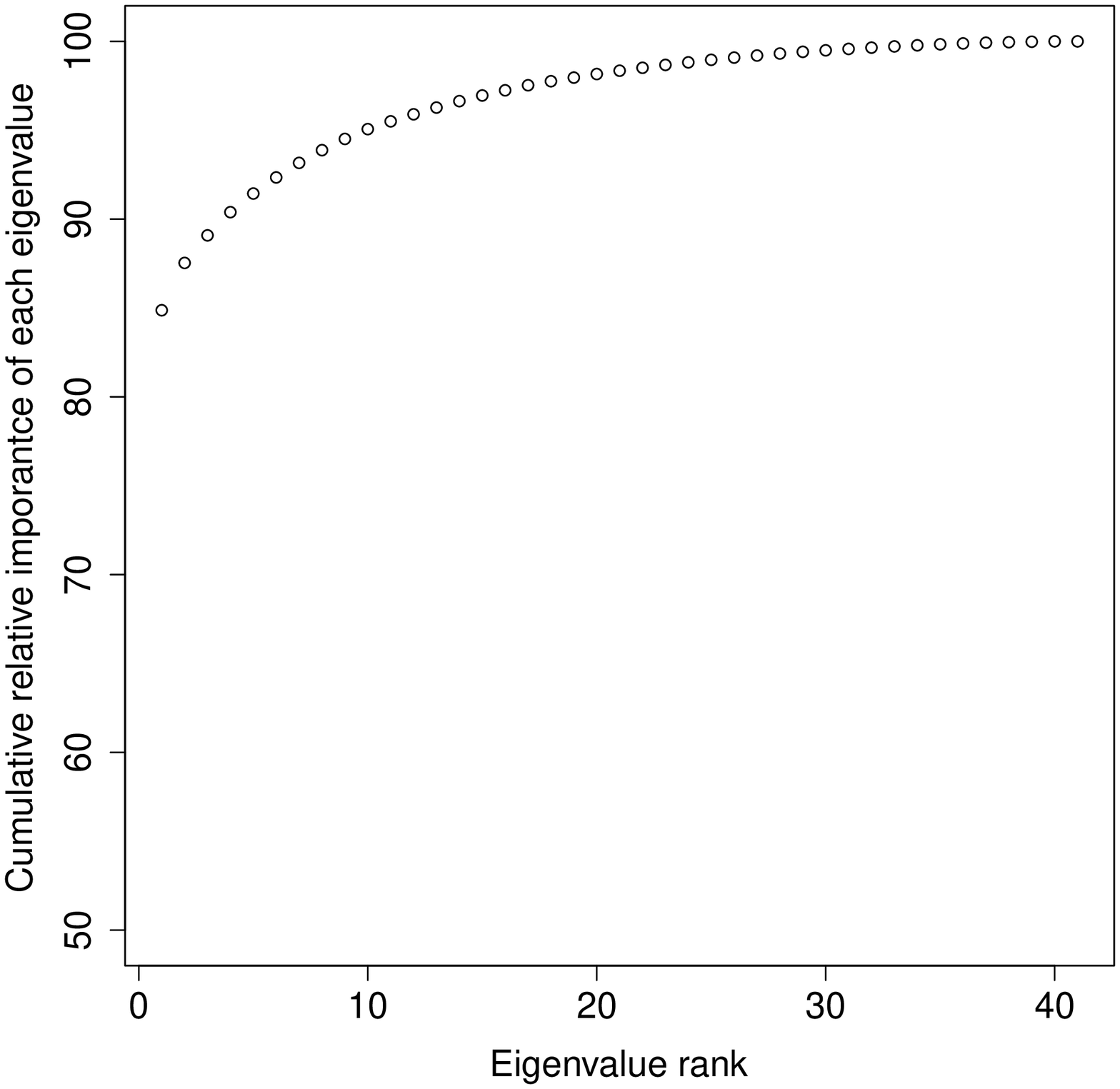}
	\caption{\textbf{PCA of the sorter data:}  On the left we see the PCA plot of the sorters.  PC1 is the coordinate along the first principal axis, and PC2 is the coordinate along the second principal axis.  Sorter known by us to be experts are marked by circles while sorters known by us to be novices are marked by triangles.  Each point is labeled on the left by the sorter number.  The second principal component discriminates experts from novices.  On the right is a plot of the cumulative relative importance of each subsequent principal component.  
	\label{pca}}
\end{figure*}

If this method is successful, the majority of variation can be explained with just a few components.  In our case, taking just the first two components explains approximately 87\% of the variability in our dataset. We thus focus on this reduced-dimension PCA, which can easily be visualized in Figure~\ref{pca}. We can now visualize our sorters and easily interpret what we see. The question of what sorter characteristic results in what behavior of PC1 and PC2 is lost in a 41-dimensional rotation and subsequent projection. In other words, this abstract representation of microscopic data (the distance matrix strongly depends on problem identities) does not boil down to a simple linear combination of macroscopic features.

Making sense of PC1 and PC2 is where the previous work on graph visualization (Subsection~\ref{subsec:visualize}) and analysis (Subsections~\ref{subsec:numcat} through~\ref{subsec:avepath}), combined with the interpretation of the CCM (Subsection~\ref{subsec:ccm}) comes together. We can look at the relative placement of our sorters by the PCA, visually analyze their graphs, and attempt an interpretation of the abstract sources of variation found by the PCA.  The expert and novice identity of each sorter is a variable known only to us and not a factor in determining the placement of the sorters by the PCA

Analyzing the sorters in order of increasing PC1-coordinate (Fig.~\ref{pca}, left panel) shows that this coordinate does {\it not} distinguish experts from novices. In other words, most variance in the data is not related to the expert or novice identity of the sorters. Instead, when analyzing the graphs associated with the subjects, it turns out that PC1 mostly reflects the ``stacker'' versus ``spreader'' behavior identified through our CCM (Subsection~\ref{subsec:ccm}), which is quite independent of being an expert or a novice. Based on this result, one could argue that card-{\it sorting} experiments most strongly measure how individuals {\it sort}, and may thus be more reflecting of what that individual's office or the file system of his or her personal computer looks like than whether or not he or she is a physics expert.

The expert/novice distinction only shows up in PC2. Going along the PC2-axis in the left panel of Figure~\ref{pca}, one finds more experts with a high PC2 and more novices with a low PC2. {\it Why} that is cannot be answered at this point and is the subject of further research in our group.  

\section{Conclusion}
In our endeavor to study the categorization behavior of experts and novices, we have developed a method for analyzing expert and novice categorizations.  In the process, we have gained insight into different human cognitive structures.  Rather than focusing on qualitative differences in category names, we chose to focus on the groupings of problems. In order to do this we have created a method for converting an abstract categorization into a graph, which may then be analyzed.  This conversion has laid the foundation for our method of analyzing card-sorting experiments, which is applicable in any experiment where sorters may put any single card into more than one category, a behavior which we name multiple categorization.

Using experimental data, we confirmed the null-results that experts and novices are not distinguishable based on macroscopic features of their card-sorting such as the number of categories.  This held true even when employing graph theoretical approaches. Finding these null result when comparing categorizations' macroscopic properties, we created the Cognitive Categorization Model, which provided insight into the general sorter behavior.  We found that the best fitting CCM had a multiple categorization probability that depended on the number of categories which led us to determine that sorters tended toward ``stacking'' or ``spreading'' when sorting physics problems.  A stacker tended to create a few general categories and multiply categorize more often.  A spreader tended to create many specific categories and multiply categorize less often.  This stacker vs.~spreader behavior is quite independent of the expert vs.~novice distinction between our sorters.

As macroscopic properties did not differentiate expert from novice, we studied the microscopic properties of categorizations in creating our distance metric.  This distance metric compares sorters' categorizations in a manner which takes problem identity into account.  In order to visualize the relative position of our sorters as measured by our distance metric, we employed Principal Components Analysis.  This allowed us to confirm the stacker vs.~spreader distinction as the largest source of variation among sorters.  It also fortuitously found the distinction between experts and novices as the second largest source of variation.

\section{Outlook}
In future studies, we will continue the process of making sense of the source of variation in the second principal component.  As the distance metric is sensitive to the identity of the problems, we plan to study the effects of including only a subset of problems from our original categorization set on the Principal Components Analysis.  Ongoing work in our lab has shown that the expert/novice distinction is highly sensitive to the problems selected.  There are some subsets of problems where the expert/novice distinction is stark and many others where this distinction is non-existent.  In the future, we will be looking at the properties of the problems which cause this high level of differentiation with the goal of understanding what ``rigging'' must occur in order to observe this stark distinction with a high degree of certainty.

\begin{acknowledgments}
The authors would like to thank the MSU physics faculty and the introductory physics classes in the Fall 2010/Spring 2011 semester for volunteering to sort problems for our study. We would also like to thank the anonymous reviewers of this journal for their helpful input and suggestions.
\end{acknowledgments}

\appendix
\section{Categorization Model Pseudocode \label{pcode}}
The following pseudocode creates a weighted adjacency matrix for a random categorization according to our categorization model.  This matrix may then be used to create a graph.  Increasing the utility of the adjacency matrix is the fact that many graph theory statistics are calculated using the weighted adjacency matrix or the adjacency matrix (which is a boolean version of the adjacency matrix).

\begin{center}
\scriptsize \ttfamily
\begin{minipage}{3.25in}
\mycom{Pseudocode for categorization model graph creation:}\\
for each graph\\
\mytab Q = input parameter \mycom{number of questions}\\
\mytab beta = input parameter \\
\mytab Cbar = input parameter \mycom{avg. number of categories}\\
\mytab 	C = random deviation from binomial distribution \\
\mytab  Pmult = alpha$^C$ \mycom{multiple sorting probability} \\
\mycom{Create boolean T matrix; rows are questions columns are categories}\\
	\mytab Initialize T\\
	\mytab X = randomize question numbers\\
	\mytab Y = shuffle list of category numbers from 1 to C\\
\mycom{Rule \#1:  Every category must be used}\\
     \mytab for all j in 1 to C\\
     	     \mytab \mytab T(X(j), Y(j)) = 1\\
\mycom{Rule \#2:  All questions must be categorized at least once}\\
     \mytab Z = sample the list from 1 to C with replacement Q-C times\\
     \mytab for all j in 1 to (Q-C)\\
     	     \mytab \mytab T(X(C+j), Z(j)) = 1\\
\mycom{Rule \#3:  Each question may be categorized more than once}\\
	\mytab for all zero elements left in the T matrix\\
		\mytab \mytab if (random number from 0 to 1 < Pmult) T(element) = 1\\
\mycom{Convert T matrix into adjacency matrix (adj) where}\\
	\mytab adj(i,j) = T(i,) dot T(j,)\\
\end{minipage}
\end{center}

\section{Distance metric \label{proof}}
The following distance metric quantifies the number of edges that must be added or removed from a graph to make it identical to another graph:
\begin{equation}
	d_{rs} = \frac{1}{2} \sum_{i=1}^Q \sum_{j=1}^Q \left|X^r_{ij} - X^s_{ij} \right|
\end{equation}
Where $X^r_{ij}$ is the $(i,j)^{th}$ element in the weighted adjacency matrix for reviewer $r$.  The properties of a metric are as follows:
\begin{eqnarray}
	d_{rs} &\geq& 0 \nonumber\\
	d_{rs} &=& 0 \quad \Longleftrightarrow \quad X^r = X^s \nonumber\\
	d_{rs} &=& d_{sr} \nonumber\\
	d_{rt} &\leq& d_{rs} + d_{st}
\end{eqnarray}
The first property is clearly satisfied by considering that we are summing up all positive numbers.  The second condition is satisfied because the only way that $d_{rs}=0$ is if every element of each weighted adjacency matrix is identical and if both weighted adjacency matrices are identical, then $d_{rs}=0$.  The third condition is also met due to the symmetry of the absolute value:
\begin{eqnarray}
	d_{rs} &=& \frac{1}{2} \sum_{i=1}^Q \sum_{j=1}^Q \left|X^r_{ij} - X^s_{ij} \right|\nonumber\\
				 &=& \frac{1}{2} \sum_{i=1}^Q \sum_{j=1}^Q \left|X^s_{ij} - X^r_{ij} \right|\nonumber\\
				 &=& d_{sr} 
\end{eqnarray}
Finally, we will consider the last condition.  First, we will consider the definition of the metric:
\begin{equation*}
	d_{rt} = \frac{1}{2} \sum_{i=1}^Q \sum_{j=1}^Q \left|X^r_{ij} - X^t_{ij} \right|
\end{equation*}
Next we will utilize the additive identity to insert the $X^s_{ij}$ terms into the absolute value.
\begin{equation*}
	d_{rt} = \frac{1}{2} \sum_{i=1}^Q \sum_{j=1}^Q \left|X^r_{ij} - X^s_{ij} + X^s_{ij} - X^t_{ij} \right| 
\end{equation*}
Next, we continue with the triangle inequality.
\begin{equation*}
	d_{rt} \leq \frac{1}{2} \sum_{i=1}^Q \sum_{j=1}^Q \left[ \left|X^r_{ij} - X^s_{ij}\right| + \left|X^s_{ij} - X^t_{ij} \right| \right] 
\end{equation*}
Now we distribute the term in front of the sum.
\begin{equation*}
	d_{rt} \leq \left[\frac{1}{2} \sum_{i=1}^Q \sum_{j=1}^Q \left|X^r_{ij} - X^s_{ij}\right|\right] + \left[\frac{1}{2} \sum_{i=1}^Q \sum_{j=1}^Q \left|X^s_{ij} - X^t_{ij} \right| \right] 
\end{equation*}
And then we simplify using the definition of our metric.
\begin{equation}
	d_{rt} \leq d_{rs} + d_{st} 
\end{equation}
So we have shown that this is a metric.

%\bibliography{biblio}


\end{document}